\documentclass[11pt,epsf,letterpaper]{article}%
\usepackage{setspace}
\usepackage{color}
\usepackage{amsmath}
\usepackage{amsfonts}
\usepackage{subcaption}
\usepackage{comment}

\usepackage{booktabs}
\usepackage{verbatim}
\usepackage{amssymb}
\usepackage{graphicx}%
\providecommand{\U}[1]{\protect\rule{.1in}{.1in}}
\begin{document}

\title{Multi-Skyrmions on $AdS_{2}\times S_{2}$, Rational maps and Popcorn transitions}
\author{Fabrizio Canfora$^{1}$, Gianni Tallarita$^{2}$\\$^{1}$\textit{Centro de Estudios Cient\'{\i}ficos (CECS), Casilla 1469,
Valdivia, Chile.}\\$^{2}$\textit{Departamento de Ciencias, Facultad de Artes Liberales,
Universidad Adolfo Ib\'a\~nez}, \\\textit{Santiago 7941169, Chile }\\{\small canfora@cecs.cl, gianni.tallarita@uai.cl }}
\maketitle

\begin{abstract}
By combining two different techniques to construct multi-soliton
solutions of the (3+1)-dimensional Skyrme model, the generalized hedgehog and
the rational map ansatz, we find multi-Skyrmion configurations in
$AdS_{2}\times S_{2}$. We construct Skyrmionic multi-layered configurations
such that the total Baryon charge is the product of the number of kinks along
the radial $AdS_{2}$ direction and the degree of the rational map. We show
that, for fixed total Baryon charge, as one increases the charge density on
$\partial\left(  AdS_{2}\times S_{2}\right)  $, it becomes increasingly
convenient energetically to have configurations with more peaks in the radial
$AdS_{2}$ direction but a lower degree of the rational map. This has a direct
relation with the so-called holographic popcorn transitions in which, when the
charge density is high, multi-layered configurations with low charge on each
layer are favored over configurations with few layers but with higher charge
on each layer. The case in which the geometry is $M_{2}\times S_{2}$ can also be analyzed.

\end{abstract}

\section{Introduction}

One of the most intriguing theoretical results in Quantum Field Theory (QFT
henceforth) has been the realization that fermions can emerge out of a purely
Bosonic Lagrangian as solitonic excitations (for a detailed review see
\cite{spin}). The clearest demonstration of the importance of this result,
which goes far beyond theoretical physics, is given by Skyrme's theory
\cite{skyrme} which is one of the most important models of nuclear and
particles physics. The inclusion of the famous Skyrme term \cite{skyrme}
allows the existence of static soliton solutions of the pion non-linear sigma
model with finite energy, called \textit{Skyrmions} (see~ \cite{multis2}%
\ \cite{manton} \cite{shifman2} \cite{susy}) describing Fermionic degrees of
freedom (see \cite{bala2} \cite{bala0} \cite{bala1} \cite{ANW} \cite{guada}
\cite{Marmo} \cite{moduli} \cite{skyrmonopole} \cite{rational} and references
therein). Furthermore, the wide range of applications of this theory in other
areas (such as astrophysics, Bose-Einstein condensates, nematic liquids,
multi-ferric materials, chiral magnets and condensed matter physics in general
\cite{astroSkyrme} \cite{useful1}, \cite{useful2}, \cite{useful3},
\cite{useful4}, \cite{useful5}, \cite{useful6}, \cite{useful7.1},
\cite{useful7} and \cite{useful8}) is well recognized by now.

Recently, the generalized hedgehog ansatz introduced in \cite{41} \cite{46}
\cite{56} \cite{56b} \cite{58} \cite{58b} \cite{58c}\ \cite{ACZ}
\cite{CanTalSk1}\ allowed the construction of the first analytic
multi-Skyrmions at finite density as well as the first exact gravitating
smooth, regular and topologically non-trivial Skyrmions configurations.
Interestingly enough, very similar techniques also work very well in the case
of non-Abelian monopoles \cite{CanTalMon1} \cite{CanTalMon2}.

On the other hand, a very powerful technique to construct multi-Skyrmionic
configurations in unbounded regions is given by the so-called \textit{rational
map ansatz} introduced in \cite{rational} (a detailed review is \cite{manton}%
). Such ansatz replaces the usual isospin vector of the standard spherical
hedgehog ansatz with a more general rational map between two-spheres. Then,
one can minimize the energy with respect to both the Skyrmion profile and the
rational map: the resulting analytic configurations are extremely
close\footnote{For instance, in most cases the energies of the
\textquotedblleft rational map" multi-Skyrmions differ by less than 3 \% with
respect to the energies of the corresponding numerical solutions.} to the true
numerical solutions (see, for instance, \cite{rational2} \cite{rational2.5}
and references therein). Hence, the rational map ansatz is one of the most
powerful analytic tools in Skyrme theory which provided, for instance, a very
detailed explanation of the appearance of fullerene-like structures (as it was
already argued in the original references \cite{rational} \cite{rational2}
\cite{rational2.5}). Moreover, the rational map has already been tested (and
with excellent results \cite{rationalfinite}) on curved backgrounds. Despite
the fact that the rational map approach (unlike, for instance, the generalized
hedgehog ansatz) does not produce exact solutions of the full field equations \cite{Houghton:2001fe}
(although, as already mentioned, the results are very close to the exact
ones), from the field-theoretical point of view, the rational map approach to
the Skyrme model can be interpreted as a way to construct exact solutions but
in the mean field approximation.

One of the most intriguing (and not yet fully explored) issues related with
topological charges is the analysis of their role within the AdS/CFT
correspondence \cite{malda1} (for a review see \cite{maldareview}). In this
framework, the role of Noether charges of the bulk theory is quite
well-understood from the boundary point of view. On the other hand, there are
not so many examples of sensible bulk theories with explicit multi-solitonic
solutions which are under control on the boundary. At least in the probe
limit, this analysis is especially relevant in relation with the
Sakai-Sugimoto model \cite{sakai}. In particular, all holographic models of
QCD describe baryons as topological solitons in the bulk. In the
Sakai-Sugimoto model there is an identification between baryon number and
instanton charge. Therefore, the construction of solitons in curved space,
with a prescribed topological charge is a very relevant topic especially if
the background geometry has $AdS$ asymptotics. However, a bulk soliton
description within holographic QCD and in particular within the Sakai-Sugimoto
model is very difficult not only from the analytical but also from the
numerical point of view. Nevertheless, it has been argued in \cite{pop1}
\cite{pop2} \cite{pop3} that when the density of topological charge increases,
a series of phase transitions (called \textit{popcorn transitions}) may
happen. The lack of numerical computations has led to various approximate
methods being employed to describe this phase, as follows. A very interesting
approach has been to simplify the bulk models (decreasing the number of
dimensions to 2+1: see in particular \cite{BoSut} \cite{Elliot}) in order to
be able to perform a tractable numerical analysis. In this way it has been
possible to discuss in detail the analogue of the popcorn transitions. These
are phase transitions to multi-layered rings occurring as the topological
charge (and consequently the density) increases. The motivation of the
analysis carried out in this paper is to extend the above results by
generalizing them to higher dimensions using the full original Skyrme model
without losing the anaytical and numerical control achieved in \cite{BoSut}
\cite{Elliot}.

In the present paper, we will first analyze the (3+1)-dimensional Skyrme model
on $AdS_{2}\times S_{2}$ using the generalized hedgehog ansatz. This analysis
provides one with a framework to study the holographic meaning of topological
charges in $AdS_{2}\times S_{2}$ in a situation in which Skyrmions pile up in
the radial ``holographic" direction of $AdS_{2}$ (while the energy density is
uniform in the $S_{2}$\ directions).

Then, we will further generalize the results of the first part using the
rational map formalism which is very suitable on the $AdS_{2}\times S_{2}%
$\ background metric. The rational map ansatz \cite{rational} will be used to
construct multi-layered configurations of Skyrmions in such a way that the
Skyrmions can pile up not only in the radial $AdS_{2}$ direction but also
along the angular directions of $S_{2}$. In this construction, each layer is
constructed following the rational map ansatz \cite{rational} while the layers
are piled up on the top of each others according to the rules determined in
\cite{41} \cite{46} \cite{56} \cite{56b} \cite{58} \cite{58b} \cite{58c}
\cite{ACZ} \cite{CanTalSk1}. In the present framework, we observe the analogue
of popcorn transitions in which, depending on the values of the coupling
constants of the theory, for fixed total topological charge, configurations
with more layers but lower densities are energetically favored over
configurations made of a smaller number of layers but with higher densities.
\newline

This paper is organized as follows: in the second section the theoretical
basis of the paper is introduced and reviewed, in this section we present the
original model and discuss the general hedgehog ansatz applied to our
particular geometry. Section 3 is devoted to a similar analysis using the
rational map ansatz, we also comment on holographic popcorn transitions. In
the final section, some conclusions will be drawn.

\section{Generalized hedgehog}

In this section, the first tool to construct analytic multi-Skyrmionic
configurations \cite{41} \cite{46} \cite{56} \cite{56b} \cite{58} \cite{58b}
\cite{58c} \cite{ACZ} \cite{CanTalSk1} will be shortly described. The action
$S_{Sk}$ of the $SU(2)$ Skyrme system in four dimensional space-times is
\begin{align}
S_{Sk}  &  =\frac{K}{2}\int d^{4}x\sqrt{-g}\,\mathrm{Tr}\left(  \frac{1}%
{2}L^{\mu}L_{\mu}+\frac{\lambda}{16}F_{\mu\nu}F^{\mu\nu}\right)
\ \ \ K>0\ ,\ \ \ \lambda>0\ ,\label{skyrmaction}\\
L_{\mu}  &  :=U^{-1}\nabla_{\mu}U=L_{\mu}^{i}t_{i}\ ,\ \ F_{\mu\nu}:=\left[
L_{\mu},L_{\nu}\right]  \ ,\ \ \ \hbar=1\ ,\ \ \ c=1\ ,\nonumber
\end{align}
where the Planck constant and the speed of light have been set to $1$, $K$ and
$\lambda$ are the coupling constants, $\mathbf{1}_{2}$\ is the $2\times2$
identity matrix and the $t^{i}$ are the basis of the $SU(2)$ generators (where
the Latin index $i$ corresponds to the group index). The relations of the
coupling constants $K$ and $\lambda$ with the
couplings\footnote{Experimentally, $F_{\pi}=186\ \ MeV$ ,\ from this the Skyme constant is fitted to be $e=5.45$.} $F_{\pi}$
and $e$ used in \cite{ANW} read
\[
K=\frac{1}{4}F_{\pi}^{2},\quad K\lambda=\frac{1}{e^{2}}\ .
\]
The non-linear sigma model term of the Skyrme action is necessary to take into
account pions. The second term is the only covariant term leading to a
well-defined Hamiltonian formalism in time which supports the existence of
Skyrmions. In the context of
AdS/CFT, it has been shown that the Skyrme model appears in a very natural way
\cite{sakai}: we will comment more on this in the following sections.

The Skyrme field equations are
\begin{equation}
\nabla^{\mu}L_{\mu}+\frac{\lambda}{4}\nabla^{\mu}[L^{\nu},F_{\mu\nu}]=0.
\label{nonlinearsigma1}%
\end{equation}

The following standard parametrization of the $SU(2)$-valued scalar $U(x^{\mu
})$ will be adopted%
\begin{equation}
U^{\pm1}(x^{\mu})=Y^{0}(x^{\mu})\mathbf{1}_{2}\pm Y^{i}(x^{\mu})t_{i}%
\ ,\ \ \left(  Y^{0}\right)  ^{2}+Y^{i}Y_{i}=1\ . \label{standard1}%
\end{equation}

In \cite{46} \cite{58} \cite{58b} \cite{CanTalSk1}, in order to construct
multi-Skyrmionic configurations the following class of curved background has
been considered:
\begin{align}
ds^{2} &  =g_{AB}dy^{A}dy^{B}+R_{0}^{2}(d\theta^{2}+(\sin\theta)^{2}d\phi
^{2})\ ,\label{metric}\\
y^{1} &  =t\ ,\ y^{2}=r\ ,\ \ A,B=1,2\ ,\ \ g_{AB}=g_{AB}\left(  t,r\right)
\ ,\label{metric0.1}%
\end{align}
where $g_{AB}$ is a two-dimensional metric with Lorentzian signature
\footnote{The simplest choice analyzed in that references corresponds to
$g_{AB}$ being the two-dimensional Minkowski metric $\eta_{AB}$. This case has
been also considered in \cite{SkyrmeFirst} but with different motivations.}.
The space-times described by the above metric can be visualized as
``generalized cylinders" in which the transverse sections (instead of being
two-dimensional disks) are two-spheres of constant radius $R_{0}$ while the
$t-r$ geometry is defined by the two-dimensional metric $g_{AB}\left(
t,r\right)  $. In the following analysis, $R_{0}$ will play an important role
to connect the present results with the \textit{holographic popcorn
transitions} mentioned in the introduction. The reason is that $R_{0}$ can act
as a control parameter to increase/decrease the charge density on the boundary
of $AdS_{2}\times S_{2}$. In particular, once a multi-Skyrmionic configuration
with a prescribed topological charge on the boundary is constructed, it is
possible to increase the charge density on the boundary by decreasing $R_{0}$.
In this way, we will show explicitly that popcorn-like transitions do indeed
occur in our setting.

The well known result that elementary Skyrmions should be quantized as
Fermions (which originally was derived on flat spaces) has been extended to
space-times with compact orientable three-dimensional spatial sections in
\cite{curvedquantization} (the metric in Eq. (\ref{metric}), which includes
$AdS_{2}\times S_{2}$ as a particular case, belongs to this class). 

When $g_{AB}$ is the two-dimensional Minkowski metric $\eta_{AB}$, it is
possible to construct multiple-layer multi-Skyrmionic configurations in a
finite volume without any cut-off on the coordinates. The effectiveness of
this metric choice is also shown by the results in \cite{CanTalMon1}
\cite{CanTalMon2} in which it has been shown that, unlike what happens in flat
spaces, the equations for the Yang-Mills-Higgs system (in the sector with
non-vanishing non-Abelian electric and magnetic charges) possess analytic
solutions even in the case in which the Higgs coupling is non-zero.

The hedgehog ansatz suitable to describe a spherically symmetric Skyrmion
living in the background of Eq. (\ref{metric}) is \cite{58} \cite{58b}:%
\begin{equation}
Y^{0}=\cos\alpha\ ,\ \ Y^{i}=\widehat{n}^{i}\sin\alpha\ ,\ \ \ \alpha
=\alpha(y^{A})=\alpha(r,t)\ , \label{hedge1}%
\end{equation}%
\begin{equation}
\widehat{n}^{1}=\sin\theta\cos\phi\ ,\ \ \ \widehat{n}^{2}=\sin\theta\sin
\phi\ ,\ \ \ \widehat{n}^{3}=\cos\theta\ . \label{hedge2}%
\end{equation}
With the above ansatz the Skyrme field equations reduce to the following
scalar differential equation for the Skyrmion profile $\alpha$:%

\begin{equation}
F\left(  \alpha\right)  D^{2}\alpha+G\left(  \alpha\right)  \ \left(
D\alpha\right)  ^{2}-V\left(  \alpha\right)  =0\ \ , \label{hedge3}%
\end{equation}%
\begin{align}
V\left(  \alpha\right)   &  =\frac{\sin(2\alpha)}{R_{0}^{2}}\left\{
1+\lambda\Upsilon_{0}\frac{\sin^{2}\alpha}{R_{0}^{2}}\right\}  \ ,\ \ \Upsilon
_{0}=1\ ,\label{hedge3.01}\\
G\left(  \alpha\right)   &  =\lambda\frac{\sin(2\alpha)}{R_{0}^{2}%
}\ ,\ \ F\left(  \alpha\right)  =1+\frac{2\lambda}{R_{0}^{2}}\sin^{2}\alpha\ ,
\label{hedge3.02}%
\end{align}%
\[
D^{2}\alpha=\frac{1}{\sqrt{-\det g_{AB}}}\partial_{A}\left[  \sqrt{-\det
g_{AB}}g^{AB}\partial_{B}\alpha\right]  \ ,\ \left(  D\alpha\right)
^{2}=g^{AB}\left(  \partial_{A}\alpha\right)  \left(  \partial_{B}%
\alpha\right)  \ ,
\]
where $D^{2}$ is the D`Alambertian with respect to the two-dimensional metric
$g_{AB}$, the coefficient $\Upsilon_{0}$ in Eq. (\ref{hedge3.01}) is
introduced to help the comparison with the rational map ansatz. On the other
hand, the energy-density (that is, the $0-0$ component of the energy-momentum
tensor) reads%
\begin{equation}
T_{00}=K\left\{  F\left(  \alpha\right)  \left[  \left(  \partial_{t}%
\alpha\right)  ^{2}-g_{tt}\frac{\left(  D\alpha\right)  ^{2}}{2}\right]
-g_{tt}\frac{\sin^{2}\alpha}{R_{0}^{2}}\left(  1+\lambda\frac{\sin^{2}\alpha
}{2R_{0}^{2}}\right)  \right\}  \ . \label{t00}%
\end{equation}

The winding number $W$ for a static configuration of the form in Eqs.
(\ref{hedge1}) and (\ref{hedge2}) reads:%
\begin{equation}
W=-\frac{1}{24\pi^{2}}\int\epsilon^{ijk}Tr\left(  U^{-1}\partial_{i}U\right)
\left(  U^{-1}\partial_{j}U\right)  \left(  U^{-1}\partial_{k}U\right)
=-\frac{2}{\pi}\int\left(  \alpha^{\prime}\sin^{2}\alpha\right)  dx\ .
\label{winding}%
\end{equation}

\subsection{Generalized hedgehog on $AdS_{2}\times S_{2}$}

Here we will restrict the above formalism to the $AdS_{2}\times S_{2}$\ case.
In this case, the two-dimensional metric described by $g_{AB}$ is%
\begin{equation}
g_{AB}dy^{A}dy^{B}=-\left(  1+\frac{r^{2}}{l^{2}}\right)  dt^{2}+\frac
{1}{\left(  1+\frac{r^{2}}{l^{2}}\right)  }dr^{2}\,\label{ads1}%
\end{equation}
where $l$ is the two-dimensional AdS radius.

We will only consider static configurations:%
\begin{equation}
\alpha=\alpha\left(  r\right)  \ ,\label{static}%
\end{equation}
in such a way that the\ full Skyrme field equations Eq. (\ref{hedge3}) reduce
to:%
\begin{equation}
\frac{-l^{2}\left(  1+\lambda\frac{\Upsilon_{0}}{N^{2}}\frac{\sin\alpha^{2}%
}{\left(  R_{0}^{2}/N\right)  }\right)  \sin2\alpha}{(l^{2}+r^{2})\left(
\left(  R_{0}^{2}/N\right)  +\lambda(1-\cos2\alpha)\right)  }+\frac{2r}%
{l^{2}+r^{2}}\alpha^{\prime}+\frac{\lambda\sin2\alpha}{R_{0}^{2}%
/N+\lambda(1-\cos2\alpha)}\alpha^{\prime2}+\alpha^{\prime\prime}%
=0\ \ ,\label{ads2}%
\end{equation}
where in the present case in which the rational map is trivial (as we are
using just the hedgehog ansatz for the isospin functions in Eq. (\ref{hedge2}%
)) we have%
\begin{equation}
N=1\ ,\ \ \frac{\Upsilon_{0}}{N^{2}}=1\ .\label{adsrationalmap1}%
\end{equation}

The boundary conditions needed to find solutions with non-vanishing Baryon
charge $n$ are: $\alpha(0)=n\pi$ and $\alpha(\infty)=0$. We find solutions to
Eq.(\ref{hedge3}) with these boundary conditions numerically. Our numerical
solver involves an pseudo-time relaxation procedure in which the $\alpha(r)$
profile is evolved from an initial seed which satisfies the required boundary
conditions. The differential operators are discretized by a finite difference
method using central differences. The accuracy of this procedure is $O\left(
10^{-3}\right)  $. The obtained solutions are shown in figures \ref{fig1} and
\ref{fig2} for Baryon charge up to $n=8$. \newline

The numerical analysis shows that there are kink-like solutions with arbitrary
Baryon charge which pile up in the $r$ direction. It is worth emphasizing that
the hyperbolic geometry of $AdS_{2}$ manifests itself in figures \ref{fig1}
and \ref{fig2}. The comparison with the results in \cite{58} (in which case
the two-dimensional geometry $g_{AB}$ was the flat Minkowski space-time)
clearly shows that in the $AdS_{2}$ case the Skyrmions prefer to pile up as
much as possible in a region of small $r$: this is a fingerprint of the
``confining effect" of the $AdS$ geometry. From the same figures (and taking
into account the results in \cite{58}), one can also see that when $r$ is
large enough ($r/l>6$ in our units) there is no longer enough volume to fit further Skyrmions. The obvious explanation is
that $g_{rr}$ (and, correspondingly, the factor $\sqrt{g_{space}}$ which
determines the volume of the $t=const$ sections of $AdS_{2}\times S_{2}$) is
very small for large $r$ while the Skyrmion has its own natural size
determined by the coupling constant of the theory. Thus, when $r$ is too
large, there simply is no space for extra Skyrmions. On the other hand, the
energy-density profile along the angular direction is trivial so that a
natural question arises: is it possible to probe the boundary of
$AdS_{2}\times S_{2}$ (namely, $%
%TCIMACRO{\U{211d} }%
%BeginExpansion
\mathbb{R}
%EndExpansion
\times S_{2}$) with more general multi-Skyrmionic configurations possessing
non-trivial energy-density profiles along the angular directions as well? The
rational map will answer this question.

\section{Rational map ansatz on curved spaces}

The rational map ansatz \cite{rational}\ achieves multi-Skyrmionic
configurations replacing the isospin vector $\widehat{n}^{i}$ in Eq.
(\ref{hedge2}) by a more general map between two spheres. The metric in Eq.
(\ref{metric}) can be analyzed using the rational map approach thanks to its
spherical symmetry. It is worth emphazising that the rational map ansatz is only an approximation (although a very valuable one). Consequently, the validity of all the results derived here is limited by the validity of the rational map ansatz itself. In particular, in order for this ansatz to work, only massless Pions should be considered. 

As usual for spherically symmetric backgrounds, the Skyrme field living in the
metric given by Eq. (\ref{metric}), will be parametrized as follows%
\begin{equation}
Y^{0}=\cos\alpha\ ,\ \ Y^{i}=\left(  n_{R}\right)  ^{i}\sin\alpha
\ ,\ \ \ \alpha=\alpha(r)\ ,\ \ \delta_{ij}\left(  n_{R}\right)  ^{i}\left(
n_{R}\right)  ^{j}=1 \label{rational1}%
\end{equation}%
\begin{align}
\left(  n_{R}\right)  ^{1}  &  =\frac{R+\overline{R}}{1+\left\vert
R\right\vert ^{2}}\ ,\ \ \ \left(  n_{R}\right)  ^{2}=\frac{i\left(
R-\overline{R}\right)  }{1+\left\vert R\right\vert ^{2}}\ ,\ \ \ \left(
n_{R}\right)  ^{3}=\frac{1-\left\vert R\right\vert ^{2}}{1+\left\vert
R\right\vert ^{2}}\ ,\label{rational2}\\
R  &  =R(z)\ ,\ \ z\in%
%TCIMACRO{\U{2102} }%
%BeginExpansion
\mathbb{C}
%EndExpansion
\ \ , \label{rational3}%
\end{align}
where $z$ is a complex coordinate which, using the stereographic projection,
can be identified with the coordinates on the 2-sphere:%
\[
z=\exp\left(  i\phi\right)  \tan\left(  \frac{\theta}{2}\right)  \ .
\]
From the mathematical point of view, a rational map $R(z)$ is a holomorphic
function from $S^{2}\rightarrow S^{2}$. Generically, $R(z)$ can always be
written as%
\[
R\left(  z\right)  =\frac{p\left(  z\right)  }{q\left(  z\right)  }\ ,
\]
where $p$ and $q$ are polynomials in $z$ with no common factor. The degree $N$
of the rational map $R$ is defined as%
\begin{equation}
N=\int\frac{2idzd\overline{z}}{\left(  1+\left\vert z\right\vert ^{2}\right)
^{2}}\left(  \frac{1+\left\vert z\right\vert ^{2}}{1+\left\vert R\right\vert
^{2}}\left\vert \frac{dR}{dz}\right\vert \right)  ^{2}\ , \label{rational3.5}%
\end{equation}
namely, the integrand in Eq. (\ref{rational3.5}) is the pull-back of the area
form on the target space sphere of the rational map $R$ itself. It can be
shown that the degree $N$ is equal to:%
\[
N=\max\left(  n_{p},n_{q}\right)  \ ,
\]
where $n_{p}$ and $n_{q}$ are the degrees of $p\left(  z\right)  $ and
$q\left(  z\right)  $ respectively. The winding number $W$ of the
configuration in Eqs. (\ref{rational1}), (\ref{rational2}) and
(\ref{rational3}) (which can be identified with the baryon number $B$ even in
the case of the metric in Eq. (\ref{metric}) as shown in
\cite{curvedquantization}) reads%
\begin{equation}
W=B=-\frac{1}{24\pi^{2}}\int\epsilon^{ijk}Tr\left(  U^{-1}\partial
_{i}U\right)  \left(  U^{-1}\partial_{j}U\right)  \left(  U^{-1}\partial
_{k}U\right)  =nN\ , \label{rational4}%
\end{equation}
where%
\begin{equation}
n=\left(  -\frac{2}{\pi}\int\left(  \alpha^{\prime}\sin^{2}\alpha\right)
dx\right)  \ , \label{rational4.1}%
\end{equation}
so that the winding number is the product of the contribution coming from the
profile function $\alpha$ (more precisely, the number of kinks along the
radial direction of $AdS_{2}$ or $M_{2}$) times the degree $N$ of the rational
map $R$.

When%
\begin{equation}
R=z \label{standard}%
\end{equation}
the rational map ansatz in Eqs. (\ref{rational1}), (\ref{rational2}) and
(\ref{rational3}) reduces to the hedgehog ansatz in Eq. (\ref{metric}) and the
solution of the equation for the profile Eq. (\ref{hedge3}) provides one with
an exact solution of the full Skyrme field equations. In general, the rational
map allows one to describe more general configurations and, in particular, it
provides one with a particularly elegant explanation of the appearance of
fullerene-like structures both on flat \cite{rational} \cite{rational2}
\cite{rational2.5} and curved \cite{rationalfinite} backgrounds.

The general strategy of the rational map approach\footnote{The success of this
strategy is partially based on the fact that, in the case of the standard
hedgehog ansatz the principle of symmetric criticality holds.} is to minimize the total energy with respect to both
the soliton profile and the rational map $R$. It is worth emphasizing that the
rational map minimization procedure is a two-dimensional problem which only
depends on the geometry of the two-sphere. Consequently, the numerical results
obtained in \cite{rational} \cite{rational2} are fully applicable in all the
geometries of the form in Eq. (\ref{metric}). Such a strategy is known to
provide excellent approximations to the full numerical solutions (see for
instance, \cite{rational2} \cite{rationalfinite}). One of the main advantages
of this framework is that it disentangles the radial coordinate $r$ from the
angular coordinates. It is important to note that disentangling the radial coordinate is only an approximation using the rational map ansatz (although a remarkably good one for small winding numbers, at least in the cases discussed in the literature on the rational map ansatz, see for example \cite{rational} \cite{rational2}
\cite{rational2.5} and \cite{rationalfinite}). Consequently, one can first minimize the energy
functional with respect to the rational map (given its degree $N$). Then, one
is left with an energy functional which only depends on the profile function
so that the minimization procedure is reduced to a one-dimensional problem.
The minimization of the rational map for degrees $N\leq108$ together with the
analysis of the corresponding discrete symmetries has been already performed
in detail in the literature (see for instance, \cite{rational2}%
\ \cite{rational2.5}\ and references therein): our analysis is based on the
data of \cite{rational2}. In particular, we will use the results contained in
table 1 page 17 of \cite{rational2}. As it has been already mentioned, these
results are fully applicable here as they depend only on the angular part of
the metric (the $S_{2}$ part of the four dimensional manifold). Since the rational map works very well for massless Pions and not too large topological charges, in the following numerical analysis we have limited our attention to values of the Baryon charge $B$ less than 15 (which is a range of values for $B$ that has been well-tested in the literature on the rational map ansatz). Moreover, the rational map ansatz has been analyzed in detail on flat space-times or on curved but very regular space-times with bounded curvature. Consequently, we have chosen two background metrics ($AdS_2\times S_2$ and $M_2\times S_2$) which are products of constant curvature manifolds (in particular, in both cases analyzed here the curvature is globally bounded).

As an introduction (which nonetheless presents new results) we will first
analyze the rational map ansatz in the simpler case of $M_{2}\times S_{2}%
$\ (analyzed in \cite{58} \cite{58b} \cite{CanTalSk1}) and then we will go back to the
$AdS_{2}\times S_{2}$ case.

\subsection{Rational map on $M_{2}\times S_{2}$}

As emphasized in \cite{58} \cite{58b} \cite{CanTalSk1}, the case of
$M_{2}\times S_{2}$, which corresponds to%
\[
g_{tt}=-1\ ,\ \ g_{rr}=1\ ,\ \ g_{tr}=0\ ,\ -\frac{L}{2}\leq r\leq\frac{L}%
{2}\ ,
\]
is especially suitable to describe finite-volume effects without breaking the
spherical symmetry. This geometry describes a tube-shaped region whose
transverse sections are two-spheres $S_{2}$ instead of disks.

The boundary conditions for the profile $\alpha$ in topological sector $n$ are%
\begin{equation}
\alpha\left(  \frac{L}{2}\right)  -\alpha\left(  -\frac{L}{2}\right)
=n\pi\ ,\ \ n\in%
%TCIMACRO{\U{2124} }%
%BeginExpansion
\mathbb{Z}
%EndExpansion
\ \ \Leftrightarrow\ U\left(  -\frac{L}{2}\right)  =(-1)^{n}U\left(  \frac
{L}{2}\right)  \ . \label{rational6}%
\end{equation}
Hence, a configuration of the form in Eqs. (\ref{rational1}), (\ref{rational2}%
) and (\ref{rational3}) with a profile satisfying the above boundary
conditions corresponds to a Baryon number $B=nN$. One can think of these
configurations as multi-layered Skyrmions made up by layers generated by the
rational map ansatz such that the rule to pile up neighboring layers is
determined by the procedure introduced in \cite{58} \cite{58b}
\cite{CanTalSk1}. In the case of $M_{2}\times S_{2}$, unlike what happens in
the usual cases \cite{rational} \cite{rational2} \cite{rational2.5}
\cite{rationalfinite}, the technique introduced in \cite{58} \cite{58b}
\cite{CanTalSk1} allows one to find the general solution for the equation for
the Skyrmions profile, as we will shortly see. \newline

In the present case, both the total energy of the system and the equation for
the soliton profile $\alpha$ are simple generalizations of Eq. (\ref{hedge3})
and (\ref{t00}). By direct computation, one can check that the total energy
is
\begin{align}
E  &  =\frac{K}{2}\int\left[  F_{R}(\alpha)\left(  \alpha^{\prime}\right)
^{2}+2H_{R}(\alpha)\right]  dr\ ,\label{rational7}\\
\Upsilon &  =\frac{1}{4\pi}\int\frac{2idzd\overline{z}}{\left(  1+\left\vert
z\right\vert ^{2}\right)  ^{2}}\left(  \frac{1+\left\vert z\right\vert ^{2}%
}{1+\left\vert R\right\vert ^{2}}\left\vert \frac{dR}{dz}\right\vert \right)
^{4}\ . \label{rational8}%
\end{align}
where%
\begin{align}
F_{R}(\alpha)  &  =\left(  1+\frac{2\lambda}{R_{0}^{2}}N\sin^{2}\alpha\right)
\ ,\label{ra8.5}\\
H_{R}(\alpha)  &  =\frac{\sin^{2}\alpha}{R_{0}^{2}}\left(  N+\frac{\lambda
}{2R_{0}^{2}}\Upsilon\sin^{2}\alpha\right)  \ . \label{ra8.75}%
\end{align}
while, the equation for the Skyrmion profile $\alpha(r)$ reads%
\begin{equation}
F_{R}\left(  \alpha\right)  \alpha^{\prime\prime}+\frac{1}{2}\frac{d}{d\alpha
}\left(  F_{R}(\alpha)\right)  \ \left(  \alpha^{\prime}\right)  ^{2}%
-V_{R}\left(  \alpha\right)  =0\ \ , \label{rational9}%
\end{equation}%
\begin{equation}
V_{R}\left(  \alpha\right)  =\frac{N\sin(2\alpha)}{R_{0}^{2}}\left\{
1+\lambda\frac{\Upsilon}{N}\frac{\sin^{2}\alpha}{R_{0}^{2}}\right\}  \ .
\label{ra9.1}%
\end{equation}

It is interesting to compare the rational map ansatz and, in particular, Eqs.
(\ref{rational9}), (\ref{ra8.5}), (\ref{ra8.75}) and (\ref{ra9.1}) with the
corresponding equations for the standard hedgehog ansatz at finite volume in
\cite{58}. It is clear that the rational map ansatz corresponds to the
following rescalings on $R_{0}$ and $\Upsilon_{0}$ in Eqs. (\ref{hedge3}),
(\ref{hedge3.01}) and (\ref{hedge3.02}):%
\begin{equation}
R_{0}^{2}\rightarrow\overline{R}^{2}=\frac{R_{0}^{2}}{N}\ ,\ \ \ \Upsilon
_{0}\rightarrow\eta=\frac{\Upsilon}{N^{2}}\ , \label{rational00}%
\end{equation}
namely, the rational map decreases the effective area of the sections of the
tube by a factor $1/N$ (so that $\overline{R}^{2}$ in Eq. (\ref{rational00})
represents the area ``available" for elementary Skyrmions in each layer) while
the effective coupling $\eta$ induced by the rational map is the integral
$\Upsilon$ defined in Eq. (\ref{rational8}) divided by $N^{2}$.

Interestingly $\Upsilon$ grows roughly\footnote{According to all the available
numerical data (see \cite{rational2} \cite{rational2.5} and references
therein), in the large $N$ limit the effective coupling $\eta$ defined in Eq.
(\ref{rational00}) has a smooth behavior (at least until $N=108$) staying
almost constant, or slowly decreasing, when $N$ is large. The fact that
$\Upsilon$ is of order $N^{2}$ can also be understood by comparing the
definition of $N$ in Eq. (\ref{rational3.5}) with the definition of $\Upsilon
$\ in Eq. (\ref{rational8}).} as $N^{2}$ (see, for instance, the numerical
results in \cite{rational2}\ \cite{rational2.5}) and consequently $\eta$ is a
quite natural effective coupling constant. The present notations map into the
ones of \cite{rational2} as follows. The integral $\Upsilon$\ (defined in Eq.
(\ref{rational8})) in the above equations corresponds to the quantity
``$\mathit{I}$" (which appears in the third column of table 1 page 17 of
\cite{rational2}) while the degree of the rational map $N$ (defined in Eq.
(\ref{rational3.5})) corresponds to the quantity ``$B$" (which appears in the
first column of table 1 page 17 of \cite{rational2}). Therefore, the relevant
quantity $\eta$ defined in Eq. (\ref{rational00}) appears in the fourth column
of table 1 page 17 of \cite{rational2}. These values of \cite{rational2} will
be used in the numerical analysis of the following sections.

The minimization procedure for the rational map $R$ corresponds to minimizing
the integral $\Upsilon$ in Eq. (\ref{rational8}) subject to the constraint
that the degree of the rational map in Eq. (\ref{rational3.5}) is $N$ (see for
instance, \cite{rational2}\ \cite{rationalfinite} and references therein).
Hence, one can see that the effect of the rational map ansatz can be
interpreted as the rescaling in Eq. (\ref{rational00}) of the coupling
constants appearing in the field equation (\ref{hedge3}) for the profile in
the case of the standard hedgehog ansatz. Also in the case of the rational
map, the equation (\ref{rational9}) for the profile can be solved as follows%
\begin{equation}
\left(  \alpha^{\prime}\right)  ^{2}=\frac{I_{c}+2H_{R}(\alpha)}{F_{R}\left(
\alpha\right)  }\ , \label{rational10}%
\end{equation}
where $I_{c}$ is an integration constant. The dependence of the integration
constant $I_{c}$ on the number of layers $n$ is determined by the following
condition
\begin{equation}
\frac{L}{n}\overset{def}{=}l_{eff}=\pm\int_{0}^{\pi}\sqrt{\frac{F_{R}(\alpha
)}{I_{c}+2H_{R}(\alpha)}}d\alpha\ , \label{rational10.5}%
\end{equation}
where for definiteness we will consider only the $+$ sign in the above
equation and $l_{eff}$ (which is the total length of the tube divided by the
number of layers) is the effective length available for each layer along the
$r$-direction. It can be easily shown that this equation for $I_{c}$ always
has a smooth regular solution with $I_{c}\geq0$.

A non-trivial BPS bound can also be derived:%
\begin{align}
E_{tot}  &  \geq\left\vert Q_{R}\right\vert \ ,\label{rational11}\\
Q_{R}  &  =4\sqrt{2}\pi R_{0}^{2}K\int_{0}^{\pi}\left[  F_{R}(\alpha
)H_{R}(\alpha)\right]  ^{1/2}\ d\alpha\ . \label{rational12}%
\end{align}

One of the most important consequences of the fact that the equation
(\ref{rational9}) for the profile can be solved by quadrature as in Eq.
(\ref{rational10}) is that it allows one to derive a closed explicit form for
the total energy of the system as a function of the couplings of the theory
$K$ and $\lambda$, of the geometric parameters $R_{0}$ and $L$ characterizing
the tube-shaped domain, and of $n$ and $N$ which are the number of layers and
the number of particles for each layer respectively. In principle, this allows
a detailed and explicit thermodynamical analysis of the system in which the
explicit knowledge of the dependence of the Skyrmion profile $\alpha$ on $r$
is not needed at all. This can be seen as follows. One needs the explicit
dependence of the total energy on the parameters of the system which are $K$,
$\lambda$, $R_{0}$, $L$ as well as on the Baryon number and on the rational
map degree. At first glance, in order to achieve this goal one needs to know
$\alpha(r)$ and then to plug the expression into Eq. (\ref{rational7}) to
reach the desired expression for the total energy. However, a simple trick
provides us with the sought expression without the knowledge of $\alpha(r)$.
Indeed, from Eq. (\ref{rational10}) we get%
\[
dr=\left[  \frac{F_{R}\left(  \alpha\right)  }{I_{c}+2H_{R}(\alpha)}\right]
^{1/2}d\alpha
\]
and using the above expression to eliminate $dr$ from Eq. (\ref{rational7})
one obtains%
\begin{equation}
E=\frac{K}{2}\int_{0}^{n\pi}\left[  I_{c}+4H_{R}(\alpha)\right]  \left[
\frac{F_{R}\left(  \alpha\right)  }{I_{c}+2H_{R}(\alpha)}\right]
^{1/2}d\alpha\ ,\label{finalflat}%
\end{equation}
where $n$ is the number of kinks along the radial direction $r$. Thus, taking
into account that $I_{c}$ is defined in Eq. (\ref{rational10.5}), the above
equation represents an explicit expression for the total energy of the system
as a function of $K$, $\lambda$, $R_{0}$, $L$ as well as of $n$ and $N$. For
instance, one can derive the above expression with respect to $L$ and $R_{0}$
to obtain explicit expressions for the longitudinal and radial pressures
(namely, $\frac{\partial E}{\partial L}$ and $\frac{\partial E}{\partial
R_{0}}$). It is also worth mentioning that for suitable values of $I_{c}$ Eq.
(\ref{finalflat}) can be reduced to an elliptic integral \cite{Adam:2014nba}. We will not analyze
this case in detail and focus on the more interesting case of $AdS_{2}\times
S_{2}$ (in which, however, numerical analysis is needed).

\subsection{Rational map on $AdS_{2}\times S_{2}$}

\begin{figure}[ptb]
\begin{subfigure}{.5\textwidth}
\centering
\includegraphics[width=0.8\linewidth]{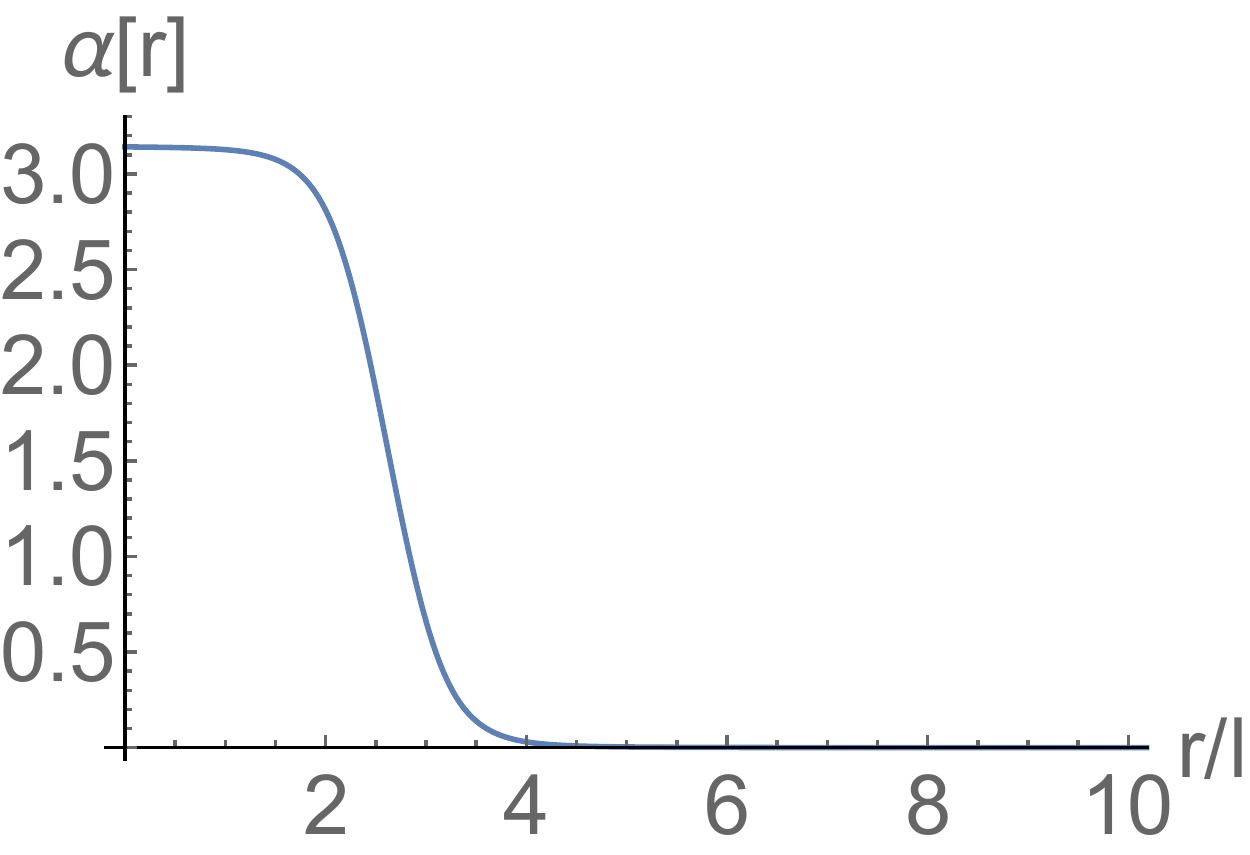}
\caption{}
\end{subfigure}
\begin{subfigure}{.5\textwidth}
\centering
\includegraphics[width=0.8\linewidth]{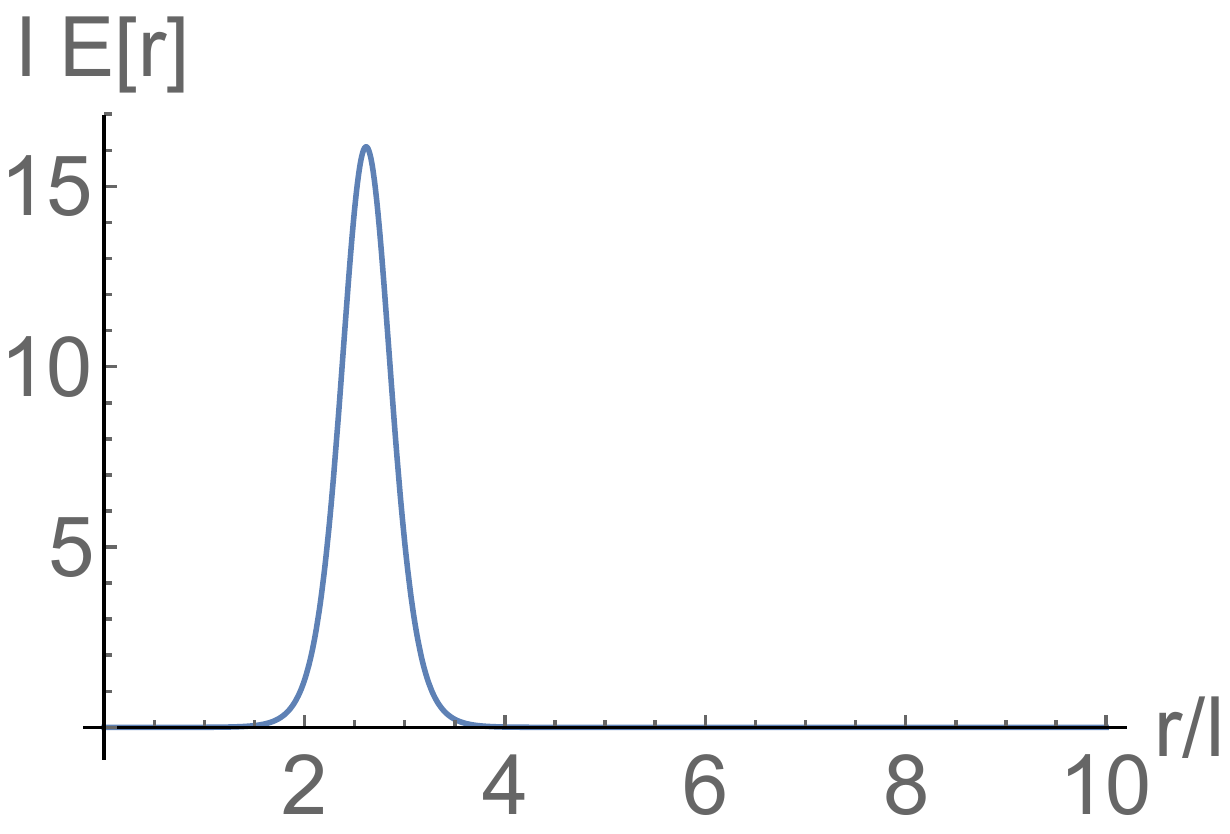}
\caption{}
\end{subfigure}
\begin{subfigure}{.5\textwidth}
\centering
\includegraphics[width=0.8\linewidth]{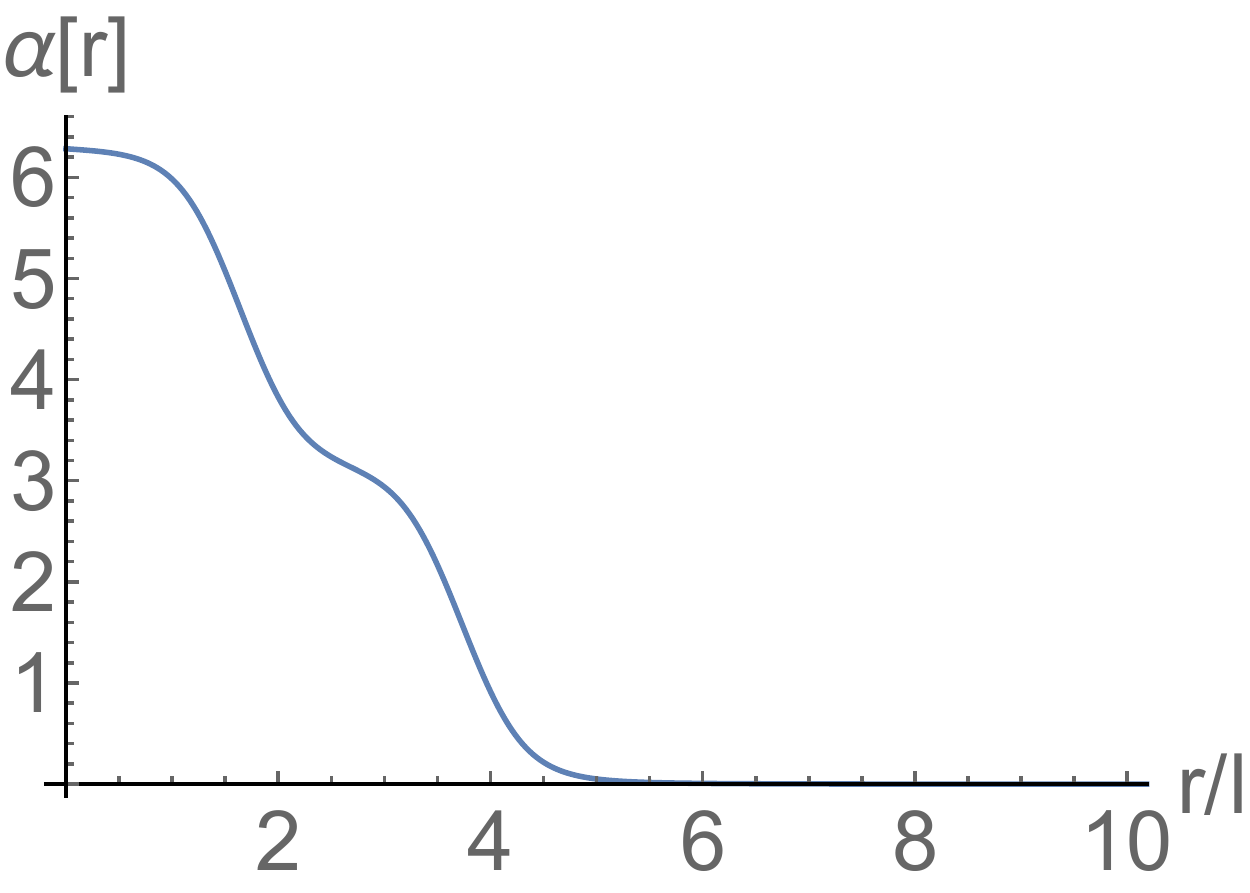}
\caption{}
\end{subfigure}
\begin{subfigure}{.5\textwidth}
\centering
\includegraphics[width=0.8\linewidth]{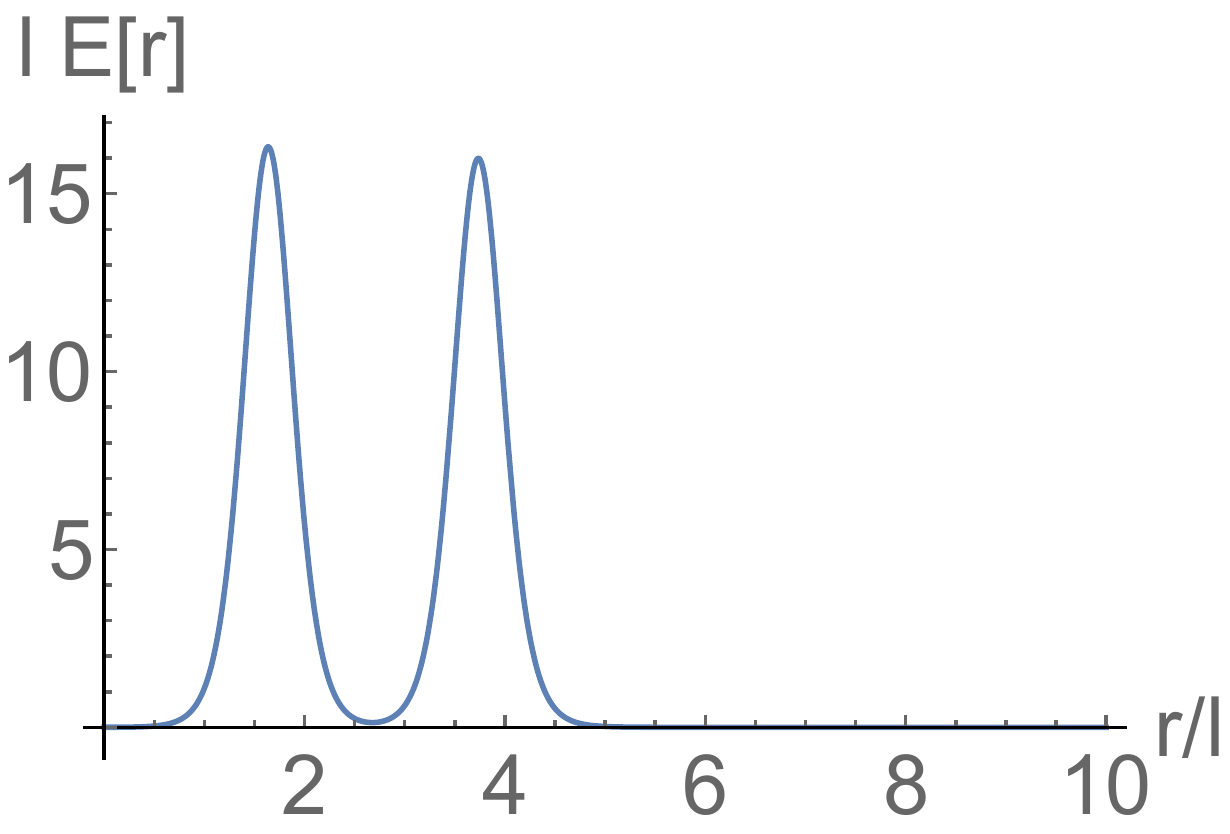}
\caption{}
\end{subfigure}
\par
\begin{subfigure}{.5\textwidth}
\centering
\includegraphics[width=0.8\linewidth]{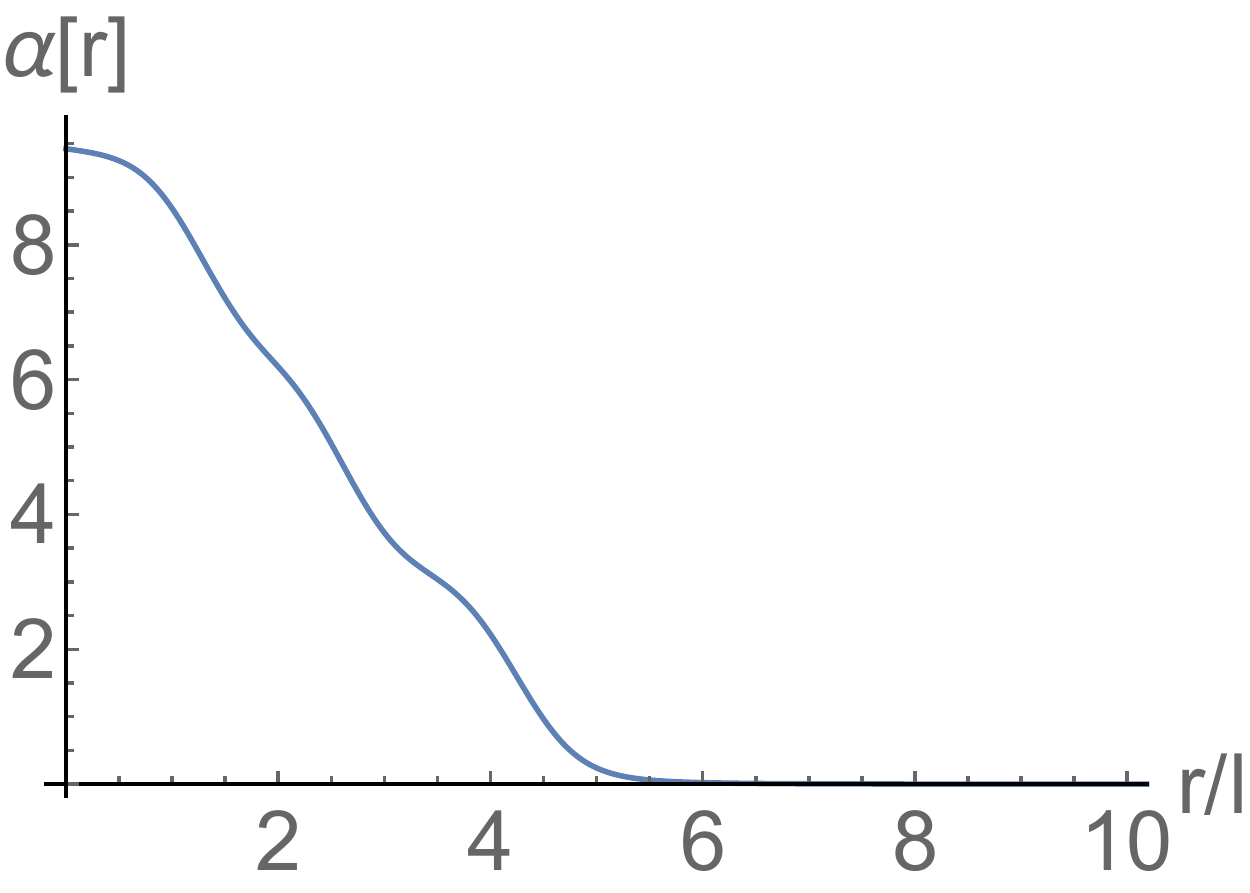}
\caption{}
\end{subfigure}
\begin{subfigure}{.5\textwidth}
\centering
\includegraphics[width=0.8\linewidth]{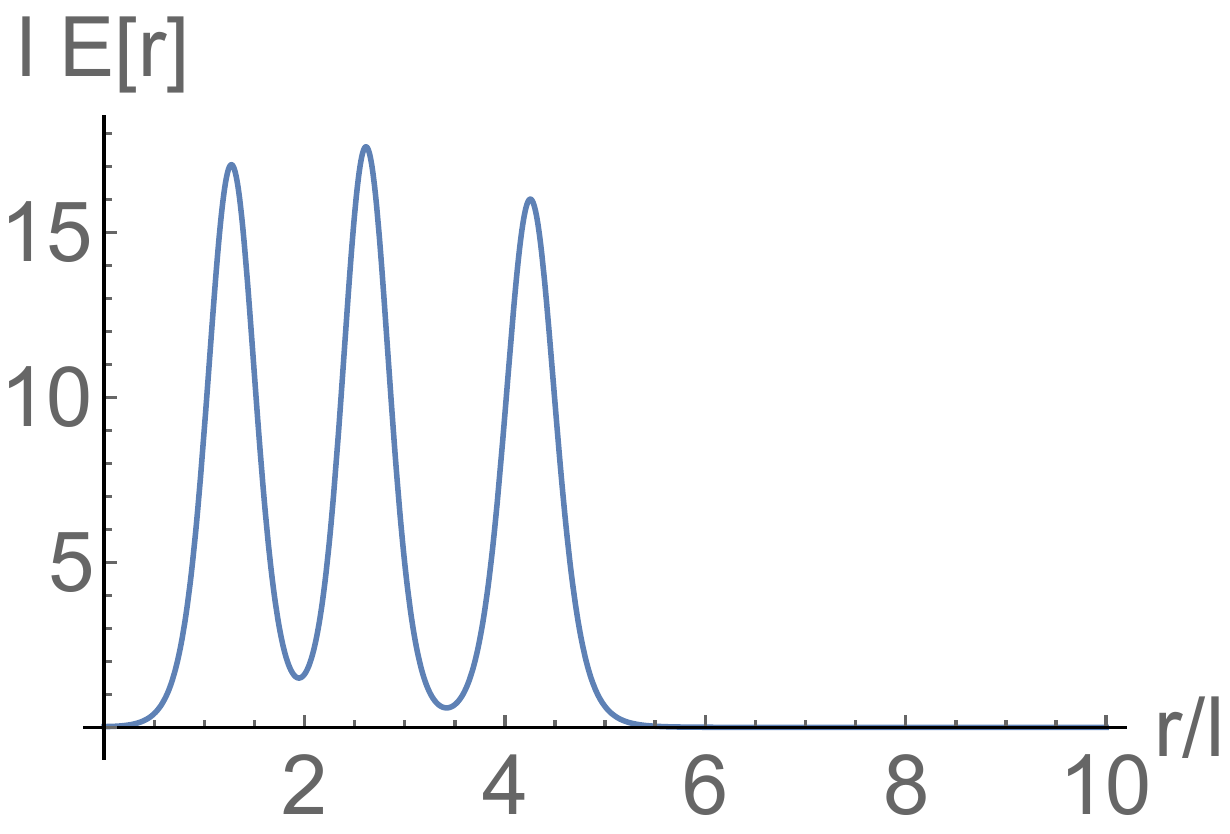}
\caption{}
\end{subfigure}
\begin{subfigure}{.5\textwidth}
\centering
\includegraphics[width=0.8\linewidth]{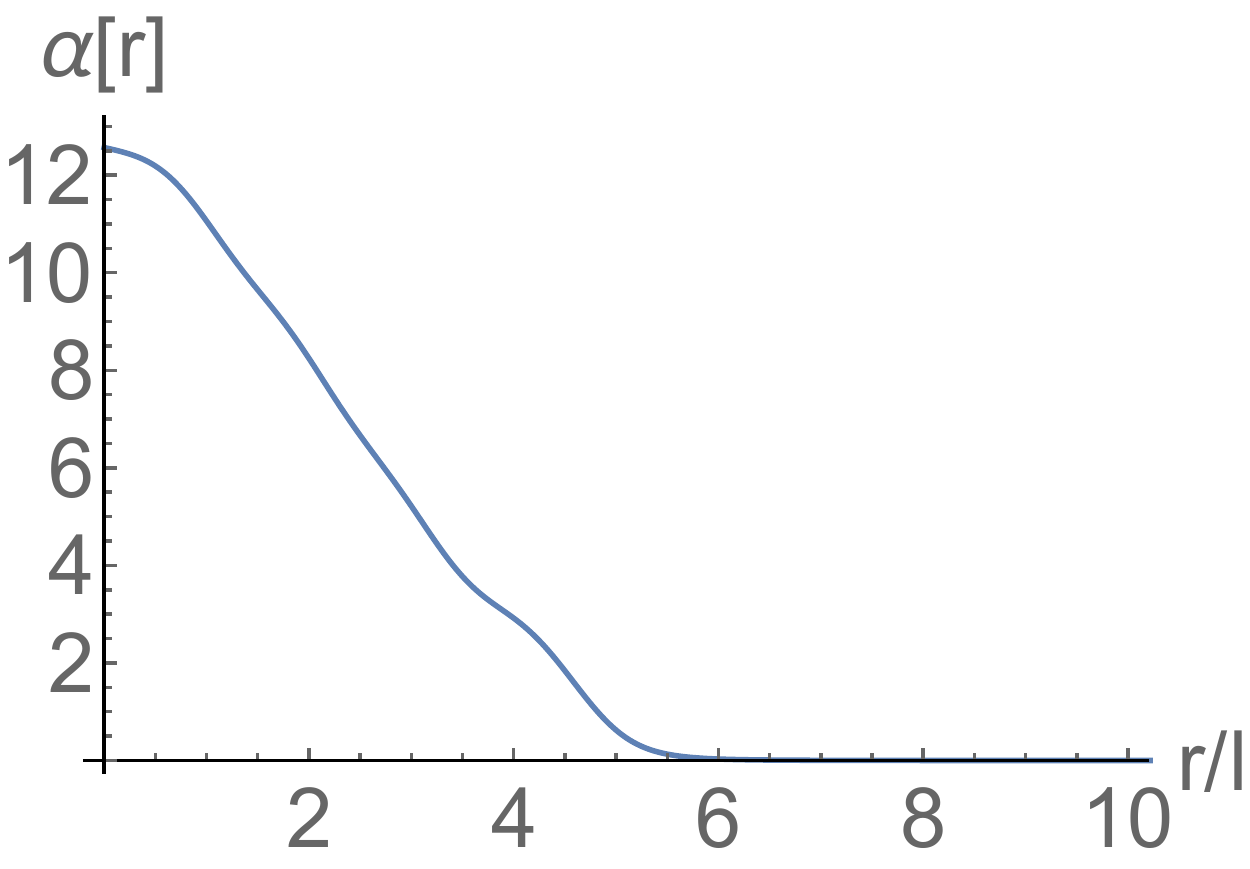}
\caption{}
\end{subfigure}
\begin{subfigure}{.5\textwidth}
\centering
\includegraphics[width=0.8\linewidth]{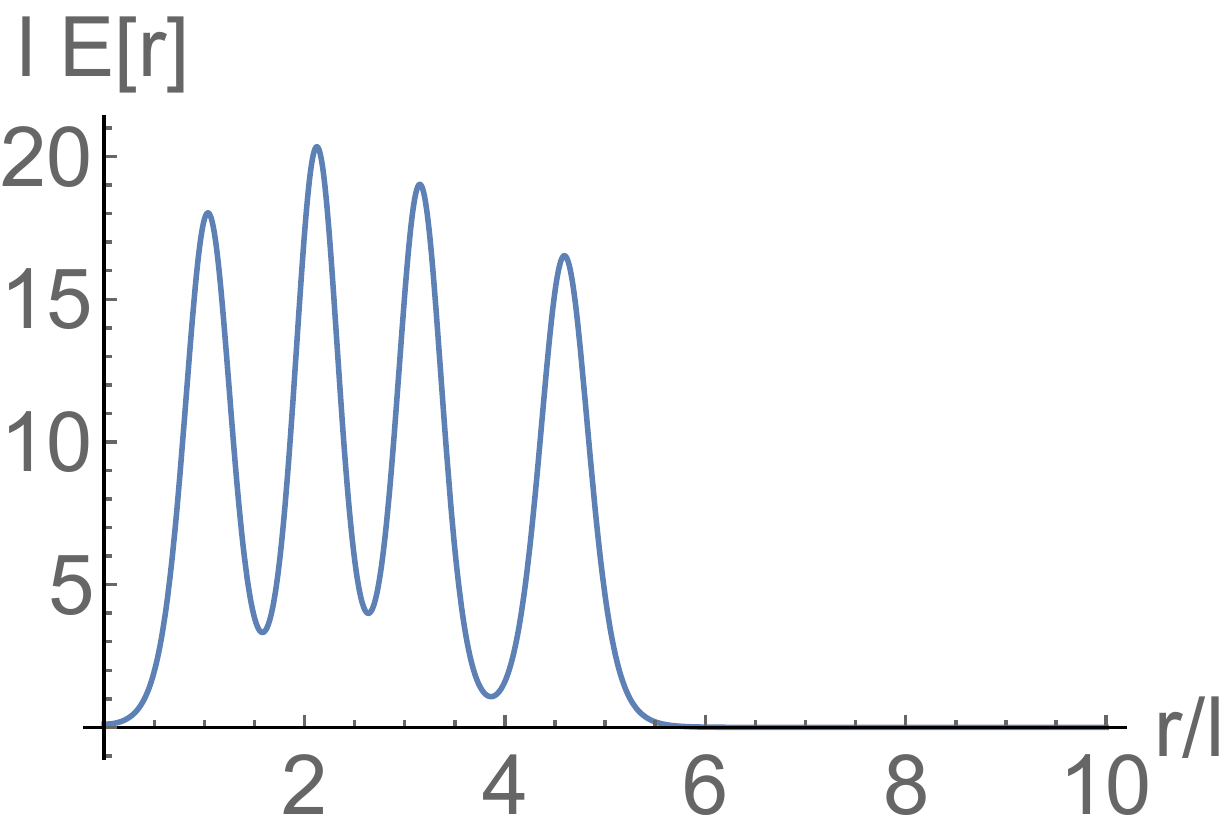}
\caption{}
\end{subfigure}
\caption{Solution profiles, alongside their respective energy profiles, at (in
units of $l=1$) $\lambda=0.1$, $R_{0}=0.4$, $N=1$, $\Upsilon=1$ and $K=1$, for
$n=1$ up to $n=4$.}%
\label{fig1}%
\end{figure}

\begin{figure}[ptb]
\begin{subfigure}{.5\textwidth}
\centering
\includegraphics[width=0.8\linewidth]{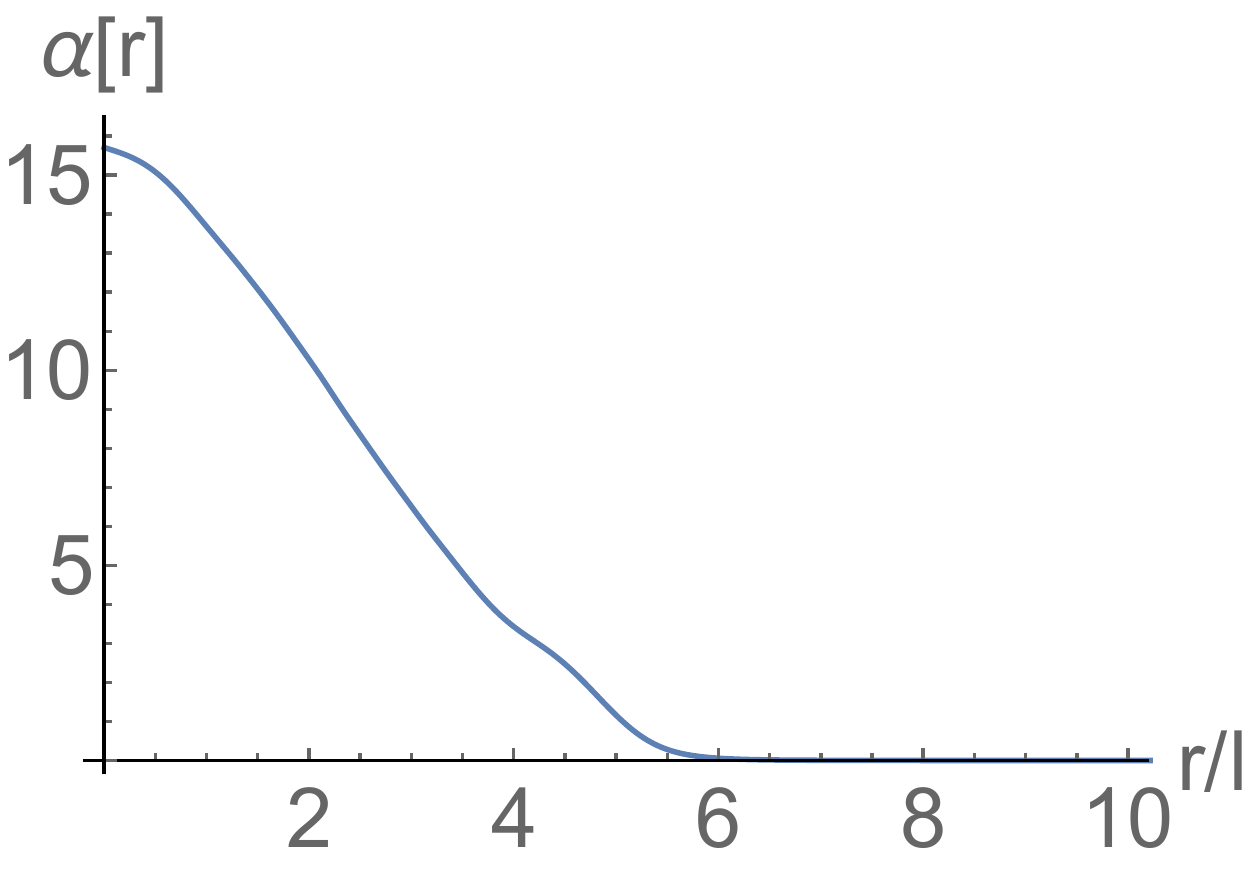}
\caption{}
\end{subfigure}
\begin{subfigure}{.5\textwidth}
\centering
\includegraphics[width=0.8\linewidth]{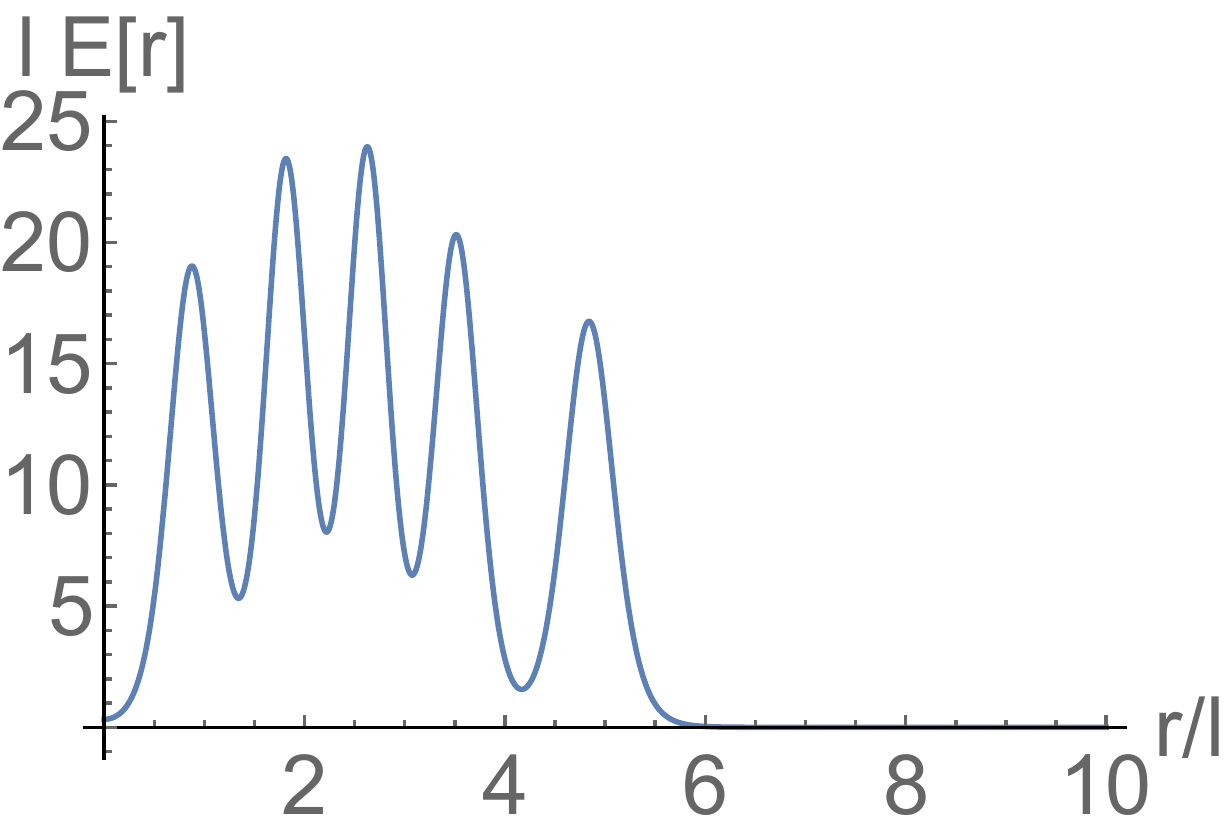}
\caption{}
\end{subfigure}
\begin{subfigure}{.5\textwidth}
\centering
\includegraphics[width=0.8\linewidth]{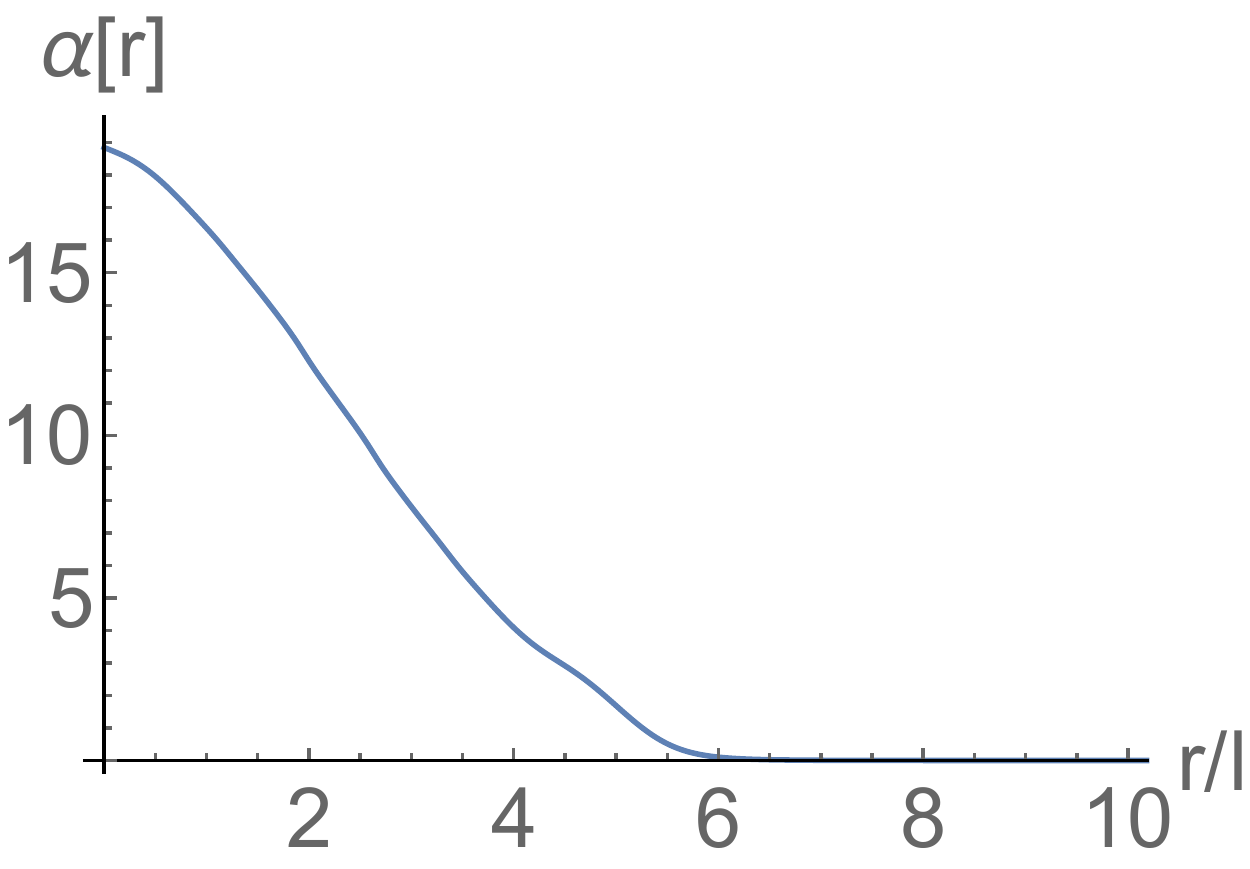}
\caption{}
\end{subfigure}
\begin{subfigure}{.5\textwidth}
\centering
\includegraphics[width=0.8\linewidth]{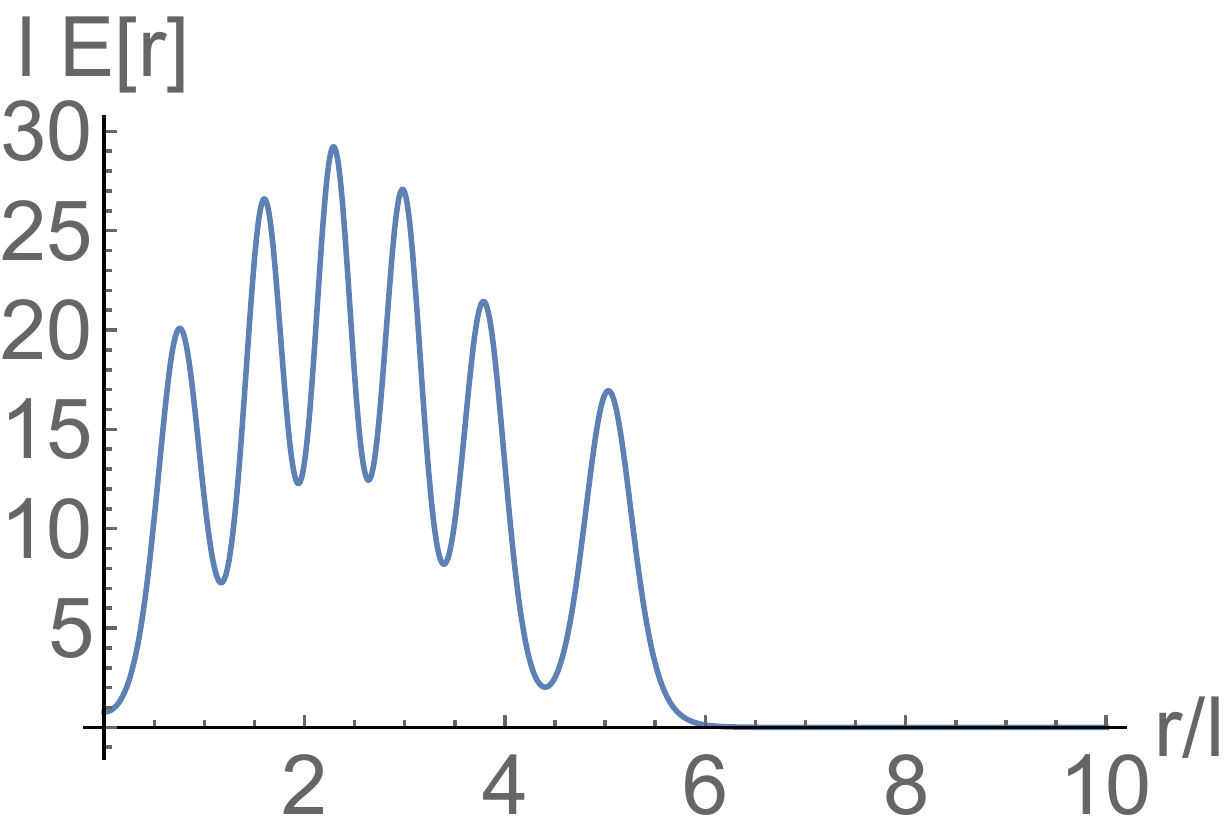}
\caption{}
\end{subfigure}
\par
\begin{subfigure}{.5\textwidth}
\centering
\includegraphics[width=0.8\linewidth]{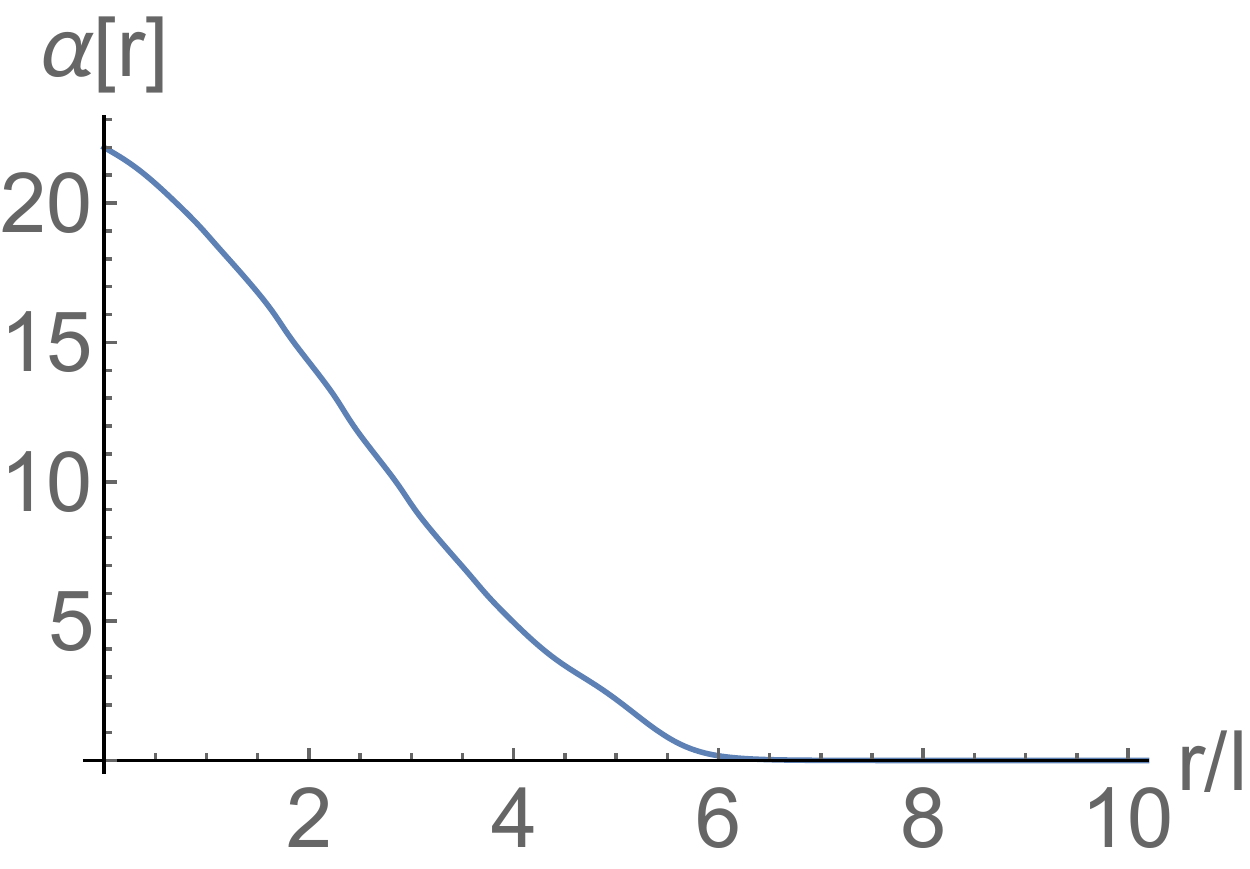}
\caption{}
\end{subfigure}
\begin{subfigure}{.5\textwidth}
\centering
\includegraphics[width=0.8\linewidth]{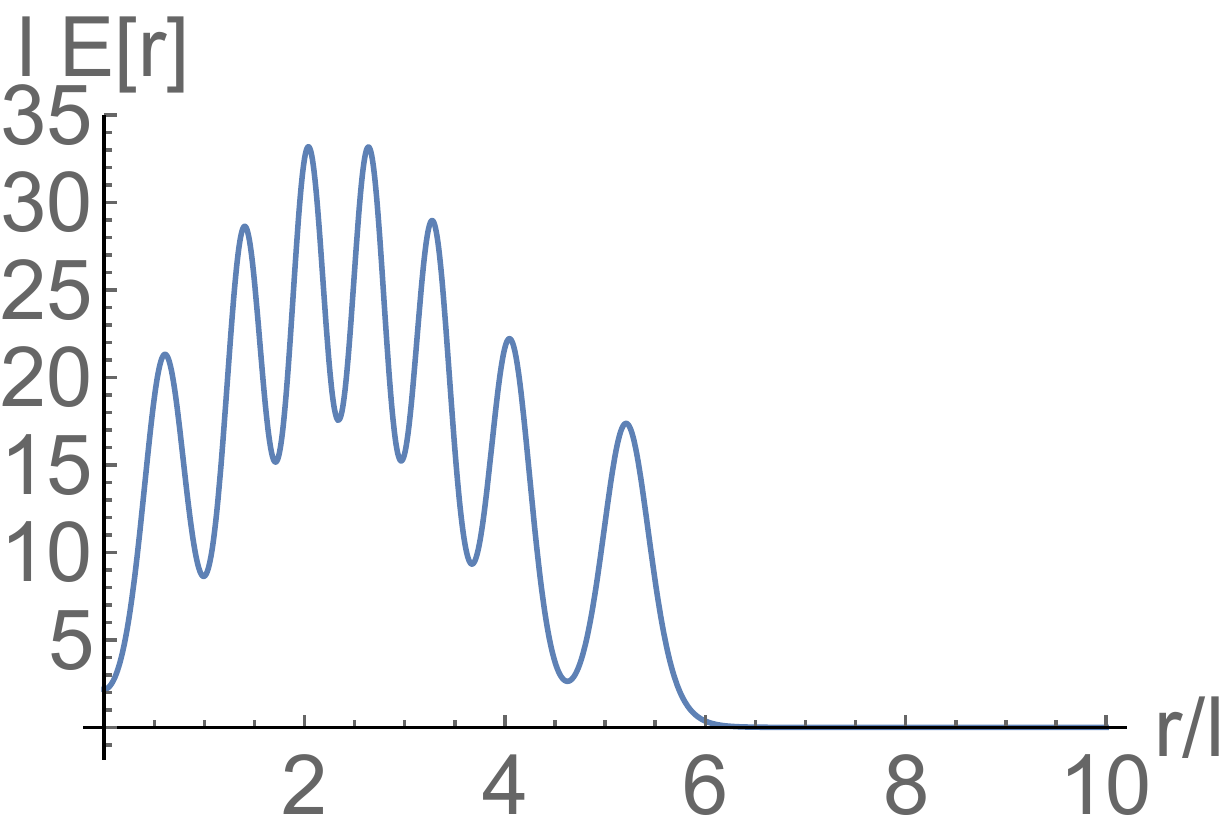}
\caption{}
\end{subfigure}
\begin{subfigure}{.5\textwidth}
\centering
\includegraphics[width=0.8\linewidth]{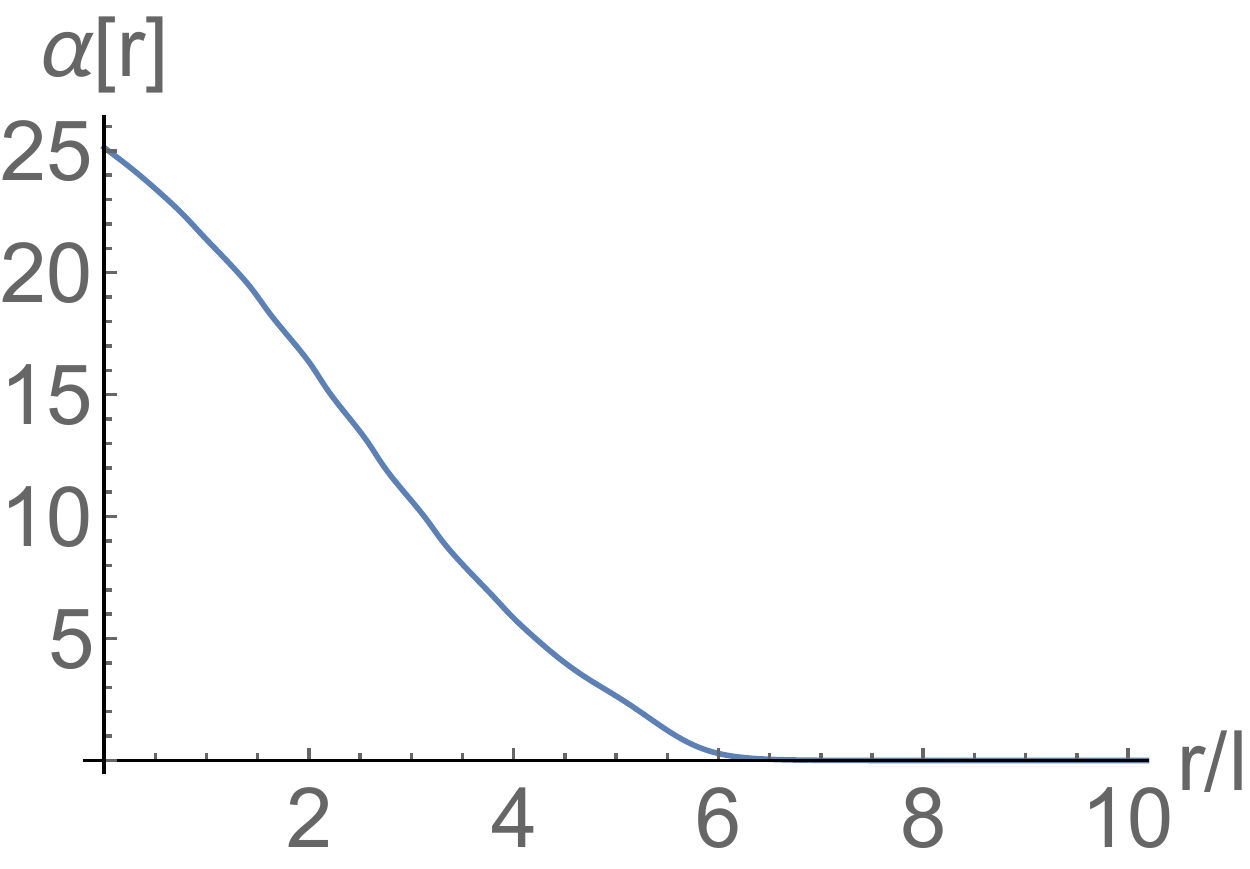}
\caption{}
\end{subfigure}
\begin{subfigure}{.5\textwidth}
\centering
\includegraphics[width=0.8\linewidth]{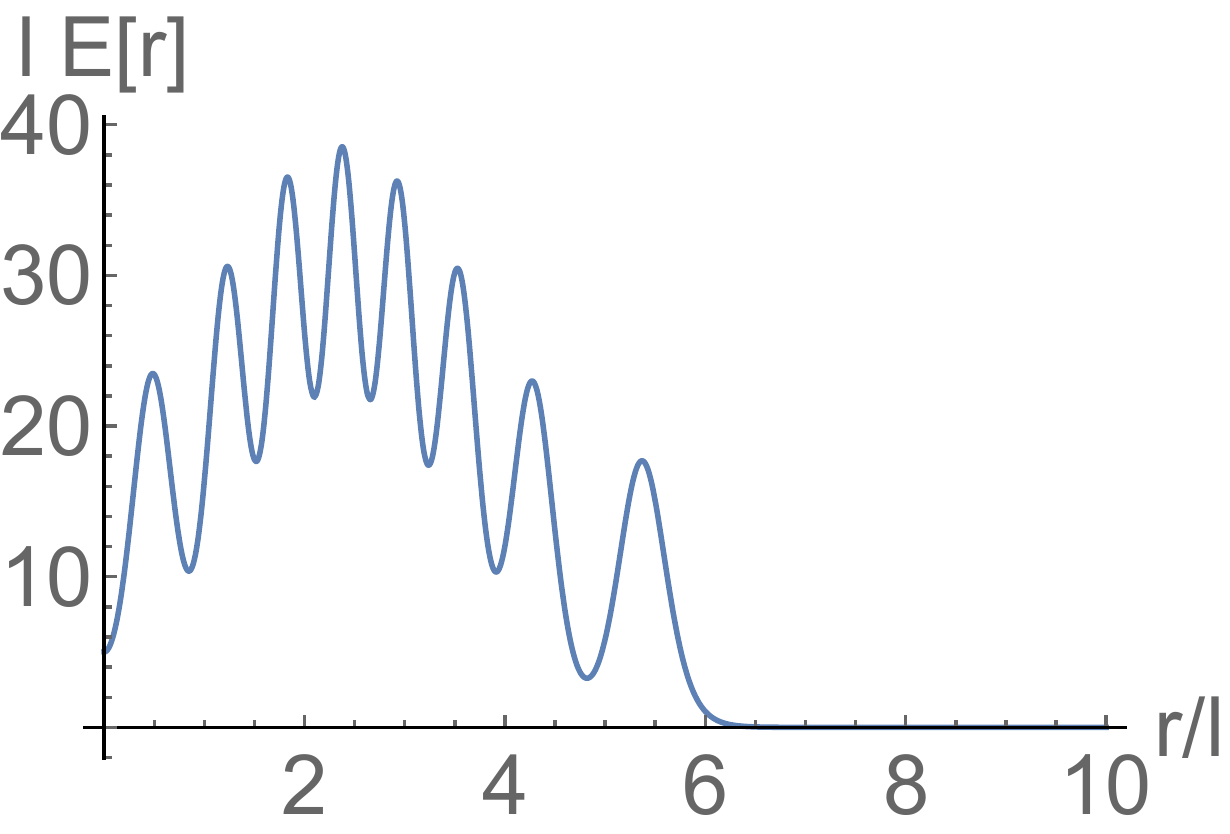}
\caption{}
\end{subfigure}
\caption{Solution profiles, alongside their respective energy profiles, at (in
units of $l=1$) $\lambda=0.1$, $R_{0}=0.4$, $N=1$, $\Upsilon=1$ and $K=1$, for
$n=5$ up to $n=8$.}%
\label{fig2}%
\end{figure}

Here we will apply the same procedure to $AdS_{2}\times S_{2}$. Namely, the
Skyrme field living in the metric in Eq. (\ref{ads1}), will be parametrized
using Eqs. (\ref{rational1}), (\ref{rational2}) and (\ref{rational3}). Then,
the second step of the method is to minimize the total energy with respect to
both the soliton profile and the rational map $R$. By doing this, one obtains
for static configurations the following equation for the profile%
\begin{equation}
F_{R}\left(  \alpha\right)  \partial_{r}\left(  \left(  1+\frac{r^{2}}{l^{2}%
}\right)  \partial_{r}\alpha\right)  +\frac{1}{2}\frac{d}{d\alpha}\left(
F_{R}(\alpha)\right)  \left(  1+\frac{r^{2}}{l^{2}}\right)  \ \left(
\alpha^{\prime}\right)  ^{2}-V_{R}\left(  \alpha\right)
=0\ \ ,\label{rationalAdS}%
\end{equation}
with the same rescaling as in Eqs. (\ref{ra8.5}), (\ref{ra9.1}) and
(\ref{rational00}). Indeed, it is easy to check that this equation can be
written in the form
\begin{equation}
\frac{-l^{2}\left(  1+\lambda\eta\frac{\sin\alpha^{2}}{\left(  R_{0}%
^{2}/N\right)  }\right)  \sin2\alpha}{(l^{2}+r^{2})\left(  \left(  R_{0}%
^{2}/N\right)  +\lambda(1-\cos2\alpha)\right)  }+\frac{2r}{l^{2}+r^{2}}%
\alpha^{\prime}+\frac{\lambda\sin2\alpha}{R_{0}^{2}/N+\lambda(1-\cos2\alpha
)}\alpha^{\prime2}+\alpha^{\prime\prime}=0\ \ ,\label{rationalADS2}%
\end{equation}
where $\eta$\ is given by Eq. (\ref{rational00}). The numerical values of the
coupling $\eta$ for many values of the degree $N$ of the rational map as well
as the corresponding discrete symmetries can, once again, be found in table 1
page 17 of \cite{rational2}. This equation should be compared to equation
(\ref{ads2}) which corresponds to the trivial rational map. We see that the
equations are extremely similar, this is why the constant $\Upsilon=1$ was
introduced previously. The numerical constants in the latter equation are
different from unity, corresponding to different degrees of the rational map.
However, the general form of the solutions will be identical, hence we refer
the reader to figures \ref{fig1} and \ref{fig2} for the general form of the
soliton profiles.

Thus, as before, we conclude that the rational map decreases the effective
area available for the Skyrmions of the ($r=const$, $t=const$) sections of
$AdS_{2}\times S_{2}$ by a factor $1/N$. On the other hand, the rational map
induces the effective coupling $\eta$ through integral $\Upsilon$ defined in
Eq. (\ref{rational8}) divided by $N^{2}$ (see, Eq. (\ref{rational00})).

\subsection{Holographic-popcorn interpretation}

The possibility to use the generalized hedgehog ansatz \cite{41} \cite{46}
\cite{56} \cite{58} \cite{58b} \cite{ACZ} \cite{CanTalSk1} together with the
rational map formalism \cite{rational} \cite{rational2} \cite{rational2.5}
\cite{rationalfinite}\ on $AdS_{2}\times S_{2}$ allows one to ask the
following natural and interesting question:

Given a set of values for the parameters of the theory (namely, $K$, $\lambda$
and $R_{0}$) and a fixed total Baryon number
\[
B=-\frac{1}{24\pi^{2}}\int\epsilon^{ijk}Tr\left(  U^{-1}\partial_{i}U\right)
\left(  U^{-1}\partial_{j}U\right)  \left(  U^{-1}\partial_{k}U\right)  =nN\
\]
(which, as explained in Eq. (\ref{rational4}), is the product of the ``number
of kinks" along the radial $AdS_{2}$ direction times the degree of the
rational map $N$), is it energetically more convenient to have higher $N$ and
lower $n$ or instead lower $N$ and higher $n$?

This question has many features in common with the holographic popcorn
transitions \cite{pop1} \cite{pop2} \cite{pop3}. In those references, it is
argued that with increasing density a series of transitions takes place where
the solitons crystal develops additional layers in the holographic direction.
A very similar phenomenon was investigated in 2+1 dimensions in \cite{BoSut}
\cite{Elliot}. In the recent publication \cite{Elliot-Ripley:2016ctk}, the author (in a low-dimensional
toy version of the Sakai-Sugimoto model) discovered the ``existence of further popcorn
transitions to three layers and beyond". It seems that these findings are quite consistent with
the present results obtained in the (3+1)-dimensional Skyrme model within the rational map
approximation in which transitions to multi-layers configurations are explicitly observed.
 This reductions in dimensions is a clear avenue to maintain
numerics under suitable control. \newline

Here we observe directly in 3+1 dimensions this kind of phenomenon in the case
in which the geometry is $AdS_{2}\times S_{2}$. A useful technical feature of
the present framework is that $R_{0}$ in Eq. (\ref{metric}) acts as a control
parameter allowing to increase/decrease the charge density on the boundary
(see the comments below Eq. (\ref{metric})). The equation for the profile (see
Eqs. (\ref{ads2}) and (\ref{rationalADS2})) produces kinks in the radial
\textquotedblleft holographic" direction. One should visualize (since the
geometry is a product $AdS_{2}\times S_{2}$) the situation as follows: at the
position $r_{i}$ of each kink (where $i=1$, ... , $n$) in the radial direction
there is a layer (extending in the $\theta$-$\varphi$ directions of $S_{2}$)
made of $N$ \textquotedblleft rational lumps". Such rational lumps (which are
piled up along the holographic radial direction) are characterized by discrete
symmetries: here we will only consider the discrete symmetries analyzed in
\cite{rational2}. In each layer, the density is $\frac{N}{R_{0}^{2}}$ so that,
for fixed $R_{0}$, increasing the degree $N$ of the rational map means
increasing the density in each layer and, in particular, on the boundary of
the manifold which is%
\[
\partial\left(  AdS_{2}\times S_{2}\right)  =%
%TCIMACRO{\U{211d} }%
%BeginExpansion
\mathbb{R}
%EndExpansion
_{t}\times S_{2}\ .
\]
Correspondingly, one expects that if the density in each layer is too high, it
is energetically favorable to decrease the well-known Skyrmion repulsion and
to have more kinks (namely, more layers) but with a lower $N$.

\begin{figure}[ptb]
\begin{subfigure}{.33\textwidth}
\centering
\begin{tabular}
[c]{|c|c|c|}\hline
nN & &$l \Delta E$\\\hline\hline\hline
2&$T_{12;21}$&2.423\\\hline\hline
4&$T_{14;41}$&8.893\\\hline
4&$T_{14;22}$&4.048\\\hline
4&$T_{41;22}$&-4.846\\\hline
6&$T_{16;61}$&20.345\\\hline
6&$T_{16;32}$&13.084\\\hline
6&$T_{16;23}$&8.541\\\hline
6&$T_{61;23}$&-11.804\\\hline
6&$T_{61;32}$&-7.261\\\hline
6&$T_{23;32}$&4.543\\\hline\hline
8&$T_{18;81}$&32.417\\\hline
8&$T_{18;42}$&22.811\\\hline
8&$T_{18;24}$&14.715\\\hline
8&$T_{81;24}$&-17.701\\\hline
8&$T_{81;42}$&-9.606\\\hline
8&$T_{24;42}$&8.095\\\hline\hline
12&$T_{1,12;12,1}$ & 62.736\\\hline
12&$T_{12,1;43}$ & -22.359\\\hline
12&$T_{12,1;34}$ & -25.416\\\hline
12&$T_{12,1;62}$ & -13.273\\\hline
12&$T_{12,1;26}$ & -39.455\\\hline
12&$T_{1,12;43}$ & 40.377\\\hline
12&$T_{1,12;34}$ & 37.320\\\hline
12&$T_{1,12;62}$ & 49.463\\\hline
12&$T_{1,12;26}$ & 23.282\\\hline
12&$T_{34;43}$ & 3.057\\\hline
12&$T_{34;26}$ & -14.039\\\hline
12&$T_{34;62}$ & 12.143\\\hline
12&$T_{43;26}$ & -17.096\\\hline
12&$T_{43;62}$ & 9.086\\\hline
12&$T_{26;62}$ & 26.181\\\hline
\end{tabular}
\caption{$R_0 = 0.1$}%
\end{subfigure}
\begin{subfigure}{.33\textwidth}
\centering
\begin{tabular}
[c]{|c|c|c|}\hline
nN && $l \Delta E$\\\hline\hline\hline
2&$T_{12;21}$&-0.063\\\hline\hline
4&$T_{14;41}$&-0.355\\\hline
4&$T_{14;22}$&0.162\\\hline
4&$T_{41;22}$&0.518\\\hline\hline
6&$T_{16;61}$&-0.942\\\hline
6&$T_{16;32}$&1.327\\\hline
6&$T_{16;23}$&1.062\\\hline
6&$T_{61;23}$&2.004\\\hline
6&$T_{61;32}$&2.268\\\hline
6&$T_{23;32}$&0.265\\\hline\hline
8&$T_{18;81}$&-2.744\\\hline
8&$T_{18;42}$&2.494\\\hline
8&$T_{18;24}$&2.360\\\hline
8&$T_{81;24}$&5.104\\\hline
8&$T_{81;42}$&5.238\\\hline
8&$T_{24;42}$&0.134\\\hline\hline
12&$T_{1,12;12,1}$ & -10.128\\\hline
12&$T_{12,1;43}$ & 16.357\\\hline
12&$T_{12,1;34}$ & 16.548\\\hline
12&$T_{12,1;62}$ & 15.571\\\hline
12&$T_{12,1;26}$ & 14.336\\\hline
12&$T_{1,12;43}$ & 6.229\\\hline
12&$T_{1,12;34}$ & 6.420\\\hline
12&$T_{1,12;62}$ & 5.443\\\hline
12&$T_{1,12;26}$ & 4.208\\\hline
12&$T_{34;43}$ & -0.192\\\hline
12&$T_{34;26}$ & -2.213\\\hline
12&$T_{34;62}$ & -0.977\\\hline
12&$T_{43;26}$ & -2.021\\\hline
12&$T_{43;62}$ & -0.786\\\hline
12&$T_{26;62}$ & 1.235\\\hline
\end{tabular}
\caption{$R_0 = 0.4$}%
\end{subfigure}
\begin{subfigure}{.33\textwidth}
\centering
\begin{tabular}
[c]{|c|c|c|}\hline
nN & &$l \Delta E$\\\hline\hline\hline
2&$T_{12;21}$&-1.281\\\hline\hline
4&$T_{14;41}$&-6.869\\\hline
4&$T_{14;22}$&-1.375\\\hline
4&$T_{41;22}$&5.494\\\hline\hline
6&$T_{16;61}$&-16.387\\\hline
6&$T_{16;32}$&-3.108\\\hline
6&$T_{16;23}$&-1.180\\\hline
6&$T_{61;23}$&15.206\\\hline
6&$T_{61;32}$&13.279\\\hline
6&$T_{23;32}$&-1.928\\\hline\hline
8&$T_{18;81}$&-31.614\\\hline
8&$T_{18;42}$&-5.791\\\hline
8&$T_{18;24}$&-0.792\\\hline
8&$T_{81;24}$&30.822\\\hline
8&$T_{81;42}$&25.823\\\hline
8&$T_{24;42}$&-4.999\\\hline\hline
12&$T_{1,12;12,1}$ & -79.427\\\hline
12&$T_{12,1;43}$ & 74.930\\\hline
12&$T_{12,1;34}$ & 78.058\\\hline
12&$T_{12,1;62}$ & 64.905\\\hline
12&$T_{12,1;26}$ & 79.178\\\hline
12&$T_{1,12;43}$ & -4.497\\\hline
12&$T_{1,12;34}$ & -1.369\\\hline
12&$T_{1,12;62}$ & -14.522\\\hline
12&$T_{1,12;26}$ & -0.249\\\hline
12&$T_{34;43}$ & -3.127\\\hline
12&$T_{34;26}$ & 1.120\\\hline
12&$T_{34;62}$ & -13.153\\\hline
12&$T_{43;26}$ & 4.248\\\hline
12&$T_{43;62}$ & -10.025\\\hline
12&$T_{26;62}$ & -14.273\\\hline
\end{tabular}
\caption{$R_0 = 0.8$}%
\end{subfigure}

\caption{Energy differences between topological sectors $T_{abcd}$ indicating
we are taking the energy difference between the sectors with topological
charge $ab$, where $a = n$ and $b = N$, and those with charge $cd$, where once
again $c=n$ and $d=N$. We present in the tables the energy differences for
overall topological charges $2,4,6,8, 12$. All values are at (in units of $l=1$)
$\lambda=0.1$, and $K=1$. }
\label{table66}%
\end{figure}

\begin{figure}
\centering
\begin{tabular}
[c]{|c|c|c|}\hline
nN & $R_0$ & $E^*$\\\hline\hline\hline
2&$0.1$&$T_{21}$\\\hline
2&$0.4$&$T_{12}$\\\hline
2&$0.8$&$T_{12}$\\\hline\hline
4&$0.1$&$T_{41}$\\\hline
4&$0.4$&$T_{22}$\\\hline
4&$0.8$&$T_{14}$\\\hline\hline
6&$0.1$&$T_{61}$\\\hline
6&$0.4$&$T_{32}$\\\hline
6&$0.8$&$T_{16}$\\\hline\hline
8&$0.1$&$T_{81}$\\\hline
8&$0.4$&$T_{42}$\\\hline
8&$0.8$&$T_{18}$\\\hline\hline
12&$0.1$&$T_{12,1}$\\\hline
12&$0.4$&$T_{34}$\\\hline
12&$0.8$&$T_{1,12}$\\\hline
\end{tabular}
\caption{Preferred energy configurations $E^*$ per topological sector as $R_0$ is varied.}%
\label{table67}
\end{figure}

The numerical results, presented in a separate page, confirm quite clearly the
\textquotedblleft popcorn pattern" mentioned above. We present our numerical
results in the tables shown in figures \ref{table66} and \ref{table67} for some characteristic topological number
combinations. We introduce the notation $T_{abcd}$ indicating we are taking
the energy difference between the sectors with topological charge $ab$, where
$a=n$ and $b=N$, and those with the same charge $cd=ab$, where once again
$c=n$ and $d=N$. The values of the effective coupling $\eta$ in Eq.
(\ref{rationalADS2}) (together with the corresponding discrete symmetries of
the rational lumps) for each value of $N$ can be read in the fourth column of
table 1 page 17 of \cite{rational2}. The most striking display of this
behavior can be seen in figure \ref{table67}.

In the case presented in figure \ref{table66} (c) the charge density on the
boundary is 64 times smaller then in the case described in figure
\ref{table66} a). Correspondingly, as shown in \ref{table67}, one can
distinctly see that (no matter the discrete symmetry of the rational
lumps\footnote{Such discrete symmetries can be read in \cite{rational2}.}) the
most energetically convenient configuration is the one with the least number
of peaks in the radial $AdS_{2}$ direction and that the energy increases by
increasing the number of radial peaks (namely, $n$).

On the other hand, for larger charge density, one can
distinctly see that (almost always) the most energetically convenient
configuration is the one with the maximum number of allowed peaks (for given
total Baryon charge) and that the energy increases by decreasing the number of
radial peaks.

Thus, the present results show convincingly in a (3+1)-dimensional setting
that, as the charge density on the boundary increases, it becomes more and
more convenient to have multi-layered configurations in which the layers have
low charge.

\section{Conclusions}

In the present paper, multi-Skyrmions of the four-dimensional Skyrme model are
constructed on $AdS_{2}\times S_{2}$. In order to achieve this goal, two
different techniques have been combined. The first one is the generalized
hedgehog ansatz while the second one is the well known rational map ansatz.
The advantage of the present geometrical setting is that, even without going
into a lower-dimensional theory, numerical analysis of multi-solitons can be
performed (at finite density too) easily. We have shown that it is possible to
solve numerically the equation for the Skyrmion profile for any degree of the
rational map. The present results give strong evidence for the correctness of the popcorn transition pattern. However, this should ideally be checked by a full 3-D numerical calculation. Moreover, we have also discussed that case of $M_{2}\times S_{2}$. Our results compare nicely with the ones in the existing literature on the subject. In particular, in \cite{Elliot} and \cite{BoSut} the authors analyzed holographic popcorns transitions within the baby Skyrme model in a pure 3d AdS background. The low dimensionality of the model makes full numerical field computations viable. The main price to pay (besides the low dimensionality) is the fact that, instead of using the original 4d Skyrme model (related to the low energy limit of QCD), they used the 3d baby Skyrme model which is a toy model. The advantage of the approach presented here is that we can analyze the original 4d Skyrme model on a four-dimensional background with a clear holographic direction (namely, $AdS_2 \times S_2$). Our results give strong evidence for the correctness of the holographic popcorn transition. The price we have pay in turn for the gain in dimensionality and for the possibility to use the 4d Skyrme model is the rational map ansatz which, in general, only produces approximated results. 

\subsection*{Acknowledgements}

This work has been funded by the Fondecyt grants 1160137, 3140122. The Centro
de Estudios Cient\'{\i}ficos (CECs) is funded by the Chilean Government
through the Centers of Excellence Base Financing Program of Conicyt.

\end{document}